\documentclass[preprint, journal]{vgtc}              

\usepackage{balance}
\preprinttext{}
\vgtccategory{Research}

\title{RAFI - A Ray/Work Forwarding
Infrastructure for Data Parallel Multi-Node/Multi-GPU Computing}

\author{
    \authororcid{Ingo~Wald}{0000-0003-0046-713X}
    \and
    \authororcid{Serkan~Demirci}{0000-0001-8805-5310}
    \and
    \authororcid{Alper~Sahistan}{0000-0002-3480-7713}
    \and
    \authororcid{Stefan Zellmann}{0000-0003-2880-9090}
    \and
    \authororcid{Andrea Paris}{0009-0004-5498-1489}
    \and
    Patrick Moran
    \and
    \authororcid{Milan Jaros}{0000-0003-4630-5339}
    \and
    \authororcid{Tatiana von Landesberger}{0000-0002-5279-1444}
    \authororcid{U\u{g}ur~G\"{u}d\"{u}kbay}{0000-0003-2462-6959}
    \and
    \authororcid{Valerio~Pascucci}{0000-0002-8877-2042}
}

\authorfooter{
  \item Ingo Wald and Andrea Paris are with NVIDIA (e-mails:~iwald@nvidia.com, aparis@nvidia.com).
  \item Serkan Demirci and U\u{g}ur~G\"{u}d\"{u}kbay are with the Department of Computer Engineering, Bilkent University, 06800 Ankara, T\"{u}rkiye (e-mails:~serkan.demirci@bilkent.edu.tr, gudukbay@cs.bilkent.edu.tr). 
  \item Alper Sahistan and Valerio Pascucci with the SCI Institute, University of Utah, Salt Lake City, UT 84112, USA (e-mails:~asahistan@sci.utah.edu, pascucci@sci.utah.edu). 
  \item Stefan Zellmann is with the Department of Computer Science, University of Cologne, 50923 Cologne, Germany (e-mail:~zellmann@uni-koeln.de).
  \item Patrick J. Moran is with NASA Ames Research Center (e-mail:~patrick.moran@nasa.gov).
  \item Milan Jaros is with IT4Innovations, VSB -Technical University of Ostrava, 17. Listopadu 2172/15, 708 00 Ostrava-Poruba, Czech Republic (e-mail:~milan.jaros@vsb.cz).
  \item Tatiana von Landesberger is with the University of Cologne, Albertus-Magnus-Platz, 50923 Cologne, NRW, Germany (e-mail:~tlandesb@uni-koeln.de).
}

\abstract{We present \rafi, a CUDA and MPI-based software framework
  that simplifies the task of building GPU-enabled data-parallel
  software where rays or similar work items need to migrate between
  different GPUs. \rafi provides a simple interface for CUDA kernels to
  \emph{forward} such work items to other GPUs, while under the hood
  managing all the CUDA and MPI-related work required to make this
  happen.  We describe \rafi's motivation and implementation, and show
  its potential in several example applications.}


\usepackage{xspace}
\usepackage[svgnames]{xcolor}
\definecolor{commentgreen}{RGB}{2,112,10}
\definecolor{eminence}{RGB}{108,48,130}
\definecolor{weborange}{RGB}{255,165,0}
\definecolor{frenchplum}{RGB}{129,20,83}
\definecolor{dkgreen}{rgb}{0,0.6,0}
\definecolor{dkblue}{rgb}{0,0,0.6}
\definecolor{gray}{rgb}{0.5,0.5,0.5}
\definecolor{mauve}{rgb}{0.58,0,0.82}
\usepackage{listings}
\renewcommand{\manuscriptnotetxt}{}
\nocopyrightspace

\graphicspath{{figures/}}
\usepackage{mathptmx}                  
\usepackage{amssymb, amsmath}
\usepackage{relsize}
\usepackage{booktabs}
\usepackage{tabu}
\usepackage{multirow}
\usepackage{tikz}
\usepackage{stackengine}
\usepackage{algorithm}
\usepackage{algpseudocode}
\usepackage{framed}
\usepackage{makecell}
\usepackage{lscape}
\usepackage{adjustbox}

\newcommand{\code}[1]{\texttt{#1}}

\definecolor{ForestGreen}{HTML}{009B55}

\lstdefinestyle{rtKernel}
{
  language=C++,
  basicstyle=\footnotesize\ttfamily\color{black},
  commentstyle=\color{ForestGreen},
  frame=tblr,
  numbers=none,
  numbersep=5pt,
  keywordstyle=\color{blue},
  keywordstyle = [2]{\color{magenta}},
  showspaces=false,
  numberstyle=\tiny\color{teal},
  showstringspaces=false,
  stringstyle=\color{red},
  tabsize=2,
  breaklines=false,                 
  captionpos=t,
  morekeywords=[1]{float2, float3, float4, log},
  morekeywords=[2]{Ray, HitRecord},
  escapeinside={<@}{@>}
}

\makeatletter
\newcommand{\thickhline}{
    \noalign {\ifnum 0=`}\fi \hrule height 1pt
    \futurelet \reserved@a \@xhline
}
\newcolumntype{"}{@{\hskip\tabcolsep\vrule width 1pt\hskip\tabcolsep}}
\makeatother

\definecolor{darkgreen}{rgb}{0,0.5,0}
\definecolor{midgreen}{rgb}{0,0.6,0}
\definecolor{lightgreen}{rgb}{0,0.8,0}
\definecolor{darkred}{rgb}{0.6,0,0}
\definecolor{darkaquamarine}{rgb}{0.18, 0.82, 0.38}

\def\brix{\emph{BriX}\xspace}
\def\optix{\emph{OptiX}\xspace}
\def\barney{\emph{Barney}\xspace}
\def\vopat{\emph{VoPaT}\xspace}
\def\rafi{\texttt{RaFI}\xspace}
\def\rafiSchlieren{\emph{SchlieRaFI}\xspace}
\def\schlierafi{\rafiSchlieren}

\def\rafiNBody{\texttt{rafi/NBody}\xspace}
\def\rafiStreamLines{\texttt{rafi/StreamLines}\xspace}
\def\schlieren{\emph{Schlieren}\xspace}

\def\cuBQL{\emph{cuBQL}\xspace}

\newif\ifdiff

\ifdiff
\usepackage{soul}

\def\removed#1{\textrm{\color{red}\st{#1}}}
\else

\def\removed#1{}
\fi

\begin{document} 

\def\raldr{\texttt{rafi/Lander}\xspace}

\firstsection{Introduction}
\maketitle

In both graphics and more general high-performance computing (HPC), many problems require simulating physical quantities such as light, particles, and radiation. One commonality when simulating these quantities is that they often require some sort of ``ray tracing'' to determine where the simulated ray or particle will travel next. How exactly this ray tracing is affected can differ across applications (and may, in particular, differ from how graphics practitioners think about this term). However, at least as long as these operations run on a single GPU, we believe this part of the problem is reasonably well understood.

A second thing common to simulations of these quantities is that the given rays or particles can travel through space, and consequently, in successive steps will often interact with different parts of the underlying domain (e.g., a particle being bounced off a surface on one side of the domain may next interact with a surface on the exact opposite side).

As long as all of the underlying data domain (i.e., the ``scene'' in graphics terminology) fits into a single GPU, this is not an issue, and will, at most, affect SIMD efficiency and/or data access coherence. However, in HPC it is common for simulations to \emph{not} run on a single GPU, and instead to be distributed across different ranks, with each GPU having only some part of the overall domain. In traditional graphics, this scenario is considered a niche problem today because most pure rendering problems typically fit into a single GPU. In more general HPC, however,
the actual input geometry  
is often tiny relative to the domain discretization and corresponding scalar field(s) that the simulation will need to store. Ironically, the visualization of such data then becomes a data-parallel problem, too, either because the derived data is large or because the user wants to perform such visualizations \emph{in situ} while the simulation is still running~\cite{childs2020terminsitu}.

When the underlying domain is split across different GPUs, rays or particles traveling across large distances in the underlying physical world means that they often need to perform their next operation on data that is not available on the GPU they are currently on, and thus, have to be sent to another rank to continue their computation---but doing so can require non-trivial amounts of code and effort.

In this paper we describe \rafi, a software library that aims at making it \emph{easier} to write GPU-enabled codes---for both rendering \emph{and} simulation---that operate on distributed data, and that require the ``forwarding'' of whatever rays, particles, or other work items the underlying code requires. To support a wide range of applications, \rafi is templated over the types of work (e.g., rays, particles, etc.) that the user's code may need to transfer. Though we will often refer to \emph{rays} from now on, this is by no means limited to distributed rendering. \rafi is particularly intended for use on GPUs, allowing a CUDA compute kernel to call \code{rafi.emitOutgoing(ray, destinationRank)}, without having to care about how exactly \rafi will internally implement this. On the backend, \rafi will batch all such forwarding requests, perform GPU-based sorting to create coherent blocks of forwarding requests, and efficiently exchange work items using properly crafted (and RDMA-capable) MPI calls in a CUDA-aware MPI environment. We describe the \rafi framework (including both interface and implementation), and demonstrate its capabilities on five different codes, including a pre-existing data-parallel path tracer (for which the use of \rafi allowed us to get significantly simpler), a new in-situ Schlieren-visualization module for CFD data, and two non-graphical codes that use \rafi for streamline and N-body computations, respectively.

\section{Background and Related Work}
\label{sec:related}

\rafi is conceptually quite similar to the well-known Ice-T~\cite{icet} framework for sort-last rendering. Though there are plenty of differences in the details (e.g., one composites images, the other forwards individual rays or work items; one builds on OpenGL, the other on CUDA; etc.), they are very similar on a conceptual level. In both cases, they work by asking the user to write per-rank kernels that generate some local data (images in one case, ray queues in the other), then the respective library performs the magic. In both cases, the goal is to relieve the user of the nitty-gritty, such as network communication. Thus, at least for those familiar with this framework, arguably the best way to view our framework is as an ``Ice-T for rays.''

Following Molnar's famous sort-first/middle/last classification of rendering algorithms~\cite{molnar1994sortingclass}, techniques based on forwarding rays would arguably fall into the sort-middle category. For a long time, this was considered too expensive, and most data parallel rendering instead relied on sort-last, but this only really makes sense for raster-based rendering. For ray tracing, various researchers have investigated ray forwarding~\cite{pprBook,reinhard::PhD,abram2018galaxy,navratil::PhD,park2018spray}, but have largely done so only on the CPU. Specifically for \rafi, our work was motivated by the \brix framework~\cite{brix}, which forwarded rays from GPU kernels. In our evaluation, we will also integrate \rafi into \vopat~\cite{vopat}, which can best be viewed as a re-work of \brix for volume data (see Section~\ref{sec:vopat}).

In our implementation, we use CUDA~\cite{cuda} for all device kernels. This framework mainly influences how rays are placed into the output queues, but it can be easily adapted to other GPU programming paradigms, such as HIP~\cite{HIP}, OpenCL~\cite{opencl}, Slang~\cite{slang}, or Kokkos~\cite{kokkos}. For all communication, we employ the Message Passing Interface (MPI)~\cite{mpi}. Implementing \rafi on other communication frameworks should be relatively straightforward. We note that for high-performance GPU communication, it is essential to leverage high-bandwidth, RDMA-accelerated direct GPU-to-GPU communication, which allows GPUs to communicate directly with the network interface controller (NIC) without first routing through host memory. Achieving this requires properly configured ``CUDA-Aware'' MPI setups~\cite{cudaaware,cudampi}. To maximize the benefits of this technology, data must be available in a few large batches.

\section{The \rafi Ray Forwarding Infrastructure}

The ultimate goal of \rafi is to provide something similar to Ice-T for data-parallel ray-based applications. This means allowing the application to specify what it wants to happen with a given set of rays on a specific rank, as well as which (possibly new) rays it wants to send to other ranks. \rafi serves as a helper library that handles tasks such as marshaling inputs, gathering and batching results, sorting data, and managing the MPI communication necessary for exchanging rays.

To achieve this, \rafi is organized into two main components: the host-side \emph{context}, which enables the application's host-side code to interact with \rafi, and the \emph{device interface}, which allows the application's CUDA code to read incoming rays and submit outgoing ones.

\subsection{``Ray''s, Templates, and Overall Software Architecture}

\rafi was initially envisioned for data-parallel ray/path tracing, and originally defined its own ray type that applications had to use. Even without realizing that the same framework is also useful way beyond just rendering, we quickly realized that even for codes that all worked on what were logically all ``rays'', these codes still all required different data layouts and definitions of what members exactly a ``ray'' would have to carry. We therefore realized that \rafi could also be---and, in fact, \emph{needed} to be---implemented via templating, where applications can specify the type of what exactly a ray is supposed to be. Due to its origins, \rafi still refers to this template parameter as a \emph{ray} type. However, it never makes any assumptions whatsoever about what this type actually contains, except that it is a \emph{trivially copyable} struct type that it can copy, move, and transmit to other ranks. The application's CUDA kernels can then ``emit'' such items to \rafi with a request to forward it to a specific rank; and that is all \rafi needs to know.

To support this templating, \rafi is organized as a header-only library, allowing the user to instantiate \rafi for any type they want; it is possible to use multiple \rafi contexts with different types in the same program or kernel.  The current version of \rafi supports only C++ (for host code) and CUDA (for device code), but realizing the same using other GPU programming languages should be straightforward.

\subsection{Ray Queues}

The general way in which an application's CUDA kernels interact with \rafi is that they read rays from an input ray queue, and emit (possibly new, or modified) rays into an output ray queue. Each ray that gets emitted will be accompanied by an \code{int} that specifies which MPI rank this ray is supposed to be sent to (which can, of course, also be the rank it is currently on).

In order to save memory, a first design used a single ray queue that served as both in- and output; any ray that was read could be written back into the same position it was originally read from (with possibly modified member values), each with an accompanying \code{int} that specified the destination rank. In that early realization, the destination rank was pre-initialized to -1, indicating that this ray did not want to go anywhere and could get discarded going forward.

This early design worked in principle but was limited in how CUDA threads could emit rays; nor did a single array save much memory, because the sorting required during the exchange stage (see below) eventually required a second copy of this array. We therefore abandoned this design and instead store inputs and outputs in two separate arrays: kernels can read from one and write to the other. No longer being limited to emitting rays to the same position as the input also allows for treating the output arrays as an atomically growing \emph{queue} of rays, where any call to \code{emitOutgoing(ray, destinationRank)} will simply append that ray to the output array (and a corresponding destination array) by increasing an atomic counter of how many rays have already been emitted, and then writing ray and destination to their respective location in those arrays. This design offers more flexibility because CUDA kernels can both consume and emit rays in any order, and without being limited to exactly one per thread.

This design needs to track four pointers: the input array, the output array, the array storing the destination ranks, and a pointer to an atomic int used to count the number of rays emitted so far. \rafi expects the application to tell it a maximum capacity for these arrays, but will otherwise manage the arrays all by itself.

During the actual ray forwarding stage, where ranks then exchange rays, these two separate arrays will also double as separate input and output arrays for sorting, as well as for the MPI send/receive operations, i.e., what was the last CUDA kernel's output array becomes the sorting stage's input array (and vice versa); the sorting stage's output array becomes the send calls' input array (and vice versa), etc.

\subsection{Device Interface}
\label{sec:device-interface}

While \rafi does allow the application to directly access the ray arrays, it would typically interact with these through what we call the
\code{device interface}: a helper class that the host can assemble, that can then be passed to CUDA as a kernel parameter, and that provides easy-to-use device functions for interacting with these queues.

\def\OPERATION#1{\par\medskip\noindent\textbf{\texttt{#1}}\xspace}

\OPERATION{DeviceInterface<T>::numIncoming()} is available on both host and device; on the host, it can, e.g., be used to determine the CUDA launch dimension; on the device, it allows a CUDA kernel to know how much valid input work there actually is.

\OPERATION{DeviceInterface<T>::getIncoming(rayID)} reads a ray from the input queue, using the given array index. 
We would expect each CUDA thread to read exactly one corresponding ray; but this is not a requirement, and rays can be read by any thread, in any number, in any order.

\OPERATION{DeviceInterface<T>::emitOutgoing(ray, dest)} atomically appends a new ray to the output queue, with \code{dest} being
an \code{int} that specifies the MPI rank to which this ray is supposed to go. This has to be a valid MPI rank, but it is perfectly valid for a rank to send to itself, too. Rays are appended atomically, so CUDA threads can emit more than one ray; however, it is the user's responsibility to appropriately size the ray queues for all rays they want to generate. Calls to \code{emitOutgoing()} that would exceed the output queue size will simply get dropped.

\subsection{\rafi Host Context}

An application that wants to use \rafi has to start by creating what we call a \emph{host context} that initializes \rafi, and manages its internal state (i.e., the ray queues, MPI communicator, etc). To create a \rafi context, the user must specify which GPU and MPI Communicator to use. If the user wants to use only some of the MPI ranks, they can do so by creating a new communicator that includes only that subset of world ranks; however, whatever communicator a context is created over is expected to be used in a \emph{collaborative} fashion.

Once created, there are only three operations with which the application interacts with this context:

\OPERATION{HostContext<T>::resizeRayQueues(N)} 
specifies the \emph{maximum} size \rafi should allocate for input, output, and destination arrays. The number of rays an array actually holds (and how many are sorted, sent, etc.) is determined by the number of rays emitted into the output array. However, this call determines the total \emph{capacity} that the kernel is allowed to emit before an output overflow would occur.

\OPERATION{HostContext<T>::getDeviceInterface()} returns an instance of the device interface described in Section~\ref{sec:device-interface}. The resulting
type is trivially copyable and can thus be passed to any
following CUDA kernel or \optix launch.

\OPERATION{HostContext<T>::forwardRays()} then implements the sending and receiving of the rays batched up in the different ranks' output queues. In this step---which in MPI-terminology operates in a \emph{collaborative} manner---each process first barriers until all other ranks have entered that call as well, then all ranks collaboratively exchange their respective batched rays such that each emitted ray ends up exactly where the corresponding \code{emitOutgoing} call indicated it should go (Section~\ref{sec:implementation-forwarding}). Once all rays are where they are supposed to be, that function returns the total number of rays still in the system, allowing the respective ranks to decide whether they need to emit additional ray-processing kernels.

\section{Implementation}

Whereas the preceding sections have described how an application would \emph{interface} with \rafi on host and/or device, this section covers exactly how these operations are realized.

Our current implementation utilizes C++, CUDA, and MPI. Implementations on other GPU frameworks, such as \code{slang} for \code{vulkan}, or \code{OpenMP}, \code{SyCL}, and \code{HIP}, should be straightforward; however, they are currently unsupported. 

All of \rafi is freely available in source form, under a permissive Apache 2 license. It comes as a header-only library with only two header files: One describes the abstract interfaces, as CUDA inline functions for the device interface, and as an abstract virtual C++ \code{Context} class for the host interface; this header file does not contain any MPI code, and can be included in different CUDA and/or C++ compilation units. The second header file contains the implementation of the host interface. It should be included in only one of the application's source files to provide the template instantiation for the virtual host interface. Both files are templated over ray type, so the application is free to use whatever type of ``ray'' it wants. The application can also use multiple different instantiations for different forwarded types if so desired.

\subsection{Host Context and Device Interface}

For the device interface, \code{getIncoming()} and \code{numIncoming()} are trivially simple \code{inline} functions; \code{emitOutgoing()} is realized through an \code{atomicAdd()} to the atomic output counter. For the host context, context initialization and ray queue management are similarly trivial; the only operation that requires some more elaboration is the actual \code{HostContext<T>::forwardRays()}.

\subsection{MPI Ray Forwarding}
\label{sec:implementation-forwarding}

Whereas the device interface was almost trivial, the forwarding operation is not, and consists of multiple steps.

\subsubsection{Sorting by Destination}

To enable efficient RDMA transfers, we first have to arrange the data to be sent and/or received into single, consistent blocks of (GPU) data---which, in our case, means sorting rays by destination. We start by swapping the input and output queues (what was previously the output array for emitting rays is now the input array for sorting), then perform a sort from this unsorted input array to the new output array.

In theory, this could be done by using GPU-based sorting directly on the ray queues, but general sorts with custom data types are expensive.  GPU sorting works best on \code{uint32} or \code{uint64} types that are amenable to radix sorting, ideally (with current sorters) in a key-only manner.  To accommodate that, we first---assuming we have $N$ input rays---generate an array of $N$ \code{uint64}s, and launch a CUDA kernel that sets the upper 32 bits of the \code{i}'th element to the desired destination rank, and the lower 32 bits to \code{i}.  This is conceptually a \code{std::pair<uint, uint>}, just stored in a single \code{uint64}.

We then use the CUDA \code{cub} library to perform a radix-sort of this array. The result is an array where necessarily all pairs with the same destination value (the upper 32 bits) are stored in consecutive parts of this array, just as we want the output rays to be; and the lower bits of each such pair give the index of the ray this pair corresponds to.

We then launch a second CUDA kernel that, for each array index \code{outIdx}, reads the given 64-bit value created in the previous kernel, extracts that pair's ray index, reads the ray from the corresponding location in the input array, and stores that in the \code{outIdx} position. This is cheap; each ray gets read exactly once and written exactly once. The result is an output array in which all rays are sorted in the same way as the 64-bit helper array; i.e., they are sorted by the destination rank
they want to go to. We now swap the input and output arrays, so the output of the sort is the input for the subsequent sending.

\subsubsection{Exchanging Rays}

We now need to know which portions of this sorted array we need to send to each rank, and similarly, how many rays each other rank will want to send us, and where we want these to be stored.

\def\bla#1{\par\medskip\noindent\textbf{#1}} \bla{Step 1: Tallying what each rank needs to send to others:} We start 
by determining where each rank's sequence of rays starts and ends in the input array. We create an array with one begin and end value for each rank, initialize these to a sentinel value (i.e., \code{\{-1,-1\}}), and then launch a kernel that checks, for each array index \code{i}, whether the destination rank for element \code{i} is different from that of index \code{i-1} and \code{i+1},
respectively. If so, that is the beginning of that element's sequence in the input array, and that value is recorded; otherwise, this thread just exists without doing anything. Each value can only be found by one thread, so no atomic operations are required.

We read this array back to the host, fill in any gaps (some ranks may not have received any rays), and also compute how many rays want to go to each rank (as the difference of their \code{begin} and \code{end} values). For each rank $r$, we now have both the \code{send\_offset[r]} (where the rays destined for $r$ are stored in the input arrays) and a \code{send\_count[r]} that states how many are destined to go to $r$.

\bla{Step 2: Exchanging how much others want to send to given rank:} Once all ranks know the workload they need to send out, we can use a \code{MPI\_Alltoall} to exchange this between ranks: Doing so means that each rank $r$ tells any other rank $s$ how many rays it wants to send to it, and correspondingly receives how many to expect in return. We then iterate over these (very few) values, compute the prefix sum, and obtain similar \code{recv\_count[r]} and \code{recv\_offset[r]} values that indicate where to direct the upcoming MPI calls in the output array to store incoming rays.

\bla{Step 3: MPI Exchange of Rays.} Each rank $r$ now knows exactly what and how much it wants to send to any rank $s$, and what and how much to expect in return; and we have non-overlapping array regions for all these sends and receives to read from and write to. We now convert from logical count/offset values to byte count/offset values by multiplying with \code{sizeof(RayT)}, and then call \code{MPI\_Alltoallv} with input/output ray arrays, and the proper count/offset arrays. Since input and outputs are device arrays, this step requires a CUDA-aware MPI setup. MPI can then perform all these sends and receives in parallel, and with GPU-direct RDMA transfer at very high
bandwidth.

\subsubsection{Wrap-up}

Once the last MPI call returns, we know that all rays on that rank have been sent and received. We can thus get ready for the application to perform its next iterations: we swap the respective input and output array pointers (the just received rays are the application's next input array), reset the atomic out queue counter to 0, and store how many rays total we have received as the new (input) ray queue size.

Additionally, we carry out a final MPI ``reduce add'' operation on the number of rays received; the result of this reduction is the total number of rays across \emph{all} ranks' output queues. Although this total does not affect the workload for the current node in the next step, it is crucial for the application's ability to properly detect ``distributed termination''~\cite{chandy-lamport}: even if a rank does not receive any work during the current iteration, it may still be assigned more work from other ranks later on if those ranks did receive tasks. Hence, knowing whether other ranks have data is still critically important. Once this reduce call is completed, we have finished exchanging rays and can return to the application.

\section{Sample Use Cases}
To demonstrate \rafi's capabilities, we will briefly describe five distinct applications we have implemented with it. These include the most obvious use case that provided the initial motivation for creating it (a data-parallel path tracer), two implementations that are suitable for {\em in situ} use that are only loosely related to ray tracing, and two that are neither ray- nor rendering-related at all (and instead perform distributed particle advection).

\begin{figure}[ht!]
\vspace*{.5em}
\begin{center}
  \def\blubb#1{\includegraphics[width=.98\columnwidth]{#1}}
  \blubb{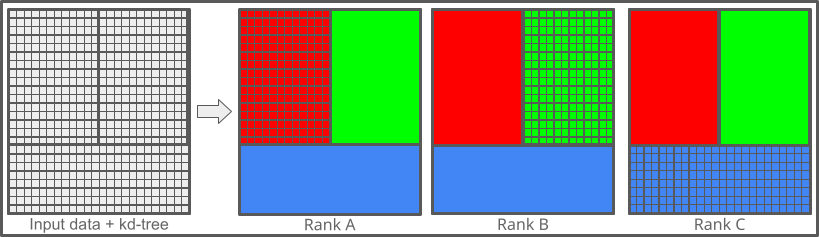}
  \\
  \blubb{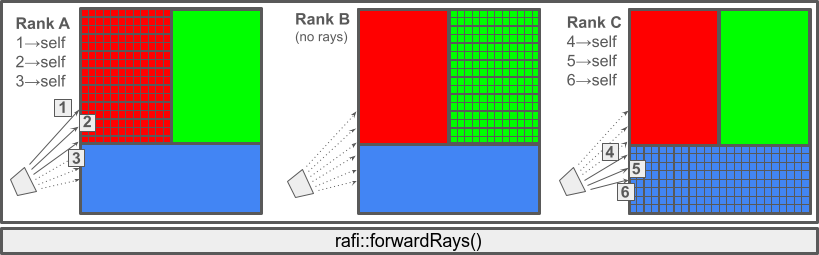}
  \\
  \blubb{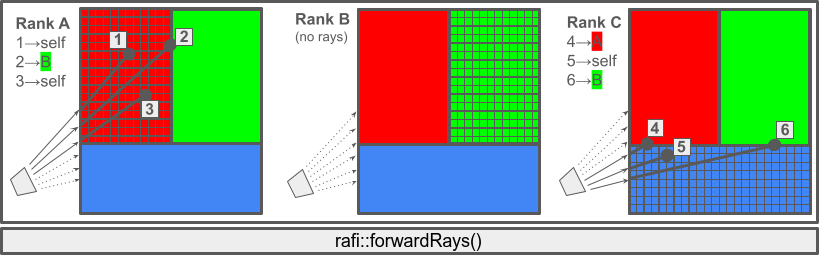}
  \\
  \blubb{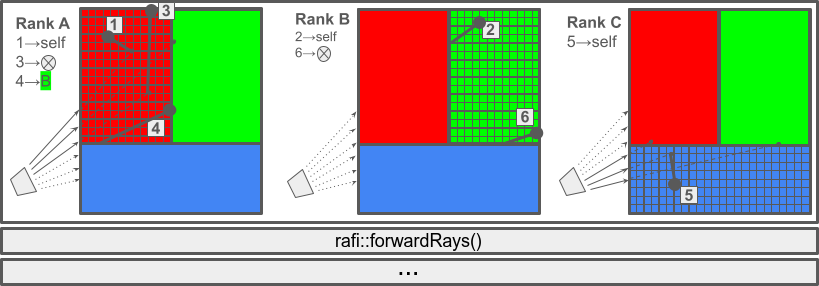}
  \\
\end{center}
\vspace*{-1.5em}
\caption{Illustration of the \vopat data-parallel volume path tracing system when using \rafi (Section~\ref{sec:vopat}). From top to bottom: 1)~Input Data: each one of three MPI-connected ranks computes the same k-d tree domain partitioning, loads its part of the input data, but also stores ``proxy'' bounding boxes that describe the other ranks' domains. 2)~RayGen: Each rank generates all primary rays and traces these rays against the proxies using OptiX. Rays hitting other ranks' proxies are discarded; those hitting the current rank are passed to \rafi to forward to itself.
3)~First actual trace: Ranks read rays from their input queue, and trace them through the local data using delta tracking. In this example, some rays (1, 3, 5) will encounter a collision in the local data---they will be scattered, then passed to \rafi for forwarding to itself in the next iteration. Other rays (2, 4, 6) pass through the local data--- they are traced against the proxy boxes to determine their destination rank, then passed to \rafi for forwarding. The \code{rafi::forwardRays()} moves rays to where they need to go, so \code{2} now ends up on rank \code{B}, \code{4} on \code{A}, and so on. From that point on, the application iterates through tracing and forwarding until all rays are processed. To avoid cluttering, we have not shown that rays can also generate additional shadow rays.\vspace*{-1em}}
\label{fig:vopat-illustration} 
\end{figure}

\subsection{VoPaT}
\label{sec:vopat}

\vopat is an experimental data-parallel path tracer for large volume data based on ray forwarding, similar to the \brix framework~\cite{brix}. \vopat was originally developed to experiment with how a \brix-like system could handle volumetric data (which \brix did not support at all); and in particular, how to realize volume
\emph{path} tracing in a data parallel manner. Like \brix, \vopat is built on the idea of \emph{ray forwarding}, where MPI-connected ranks iteratively trace wavefronts of rays through their local data, then run those rays through a next-rank kernel to determine where to go next, then send rays across ranks as required.

A step-by-step illustration of \vopat is given in Figure~\ref{fig:vopat-illustration}. Overall, there are but two different GPU kernels: one for generating an initial
ray queue of primary rays for each rank (Figure~\ref{fig:vopat-illustration}, Step 2); and 
one---the main render kernel---that reads rays from an input queue, traces them through the local data, and possibly re-emits them into an output queue for forwarding to the next step (and usually, on another rank). Once the initial ray queues are generated, rendering alternates between calling the main render kernel (to generate a new output queue) and calling the \code{vopat::forwardRays} function, which performs the actual ray movement as described above. This process continues until there are no more rays to forward. 

How exactly \vopat generates, traces, and shades rays is largely orthogonal to this paper: primary ray generation (who generates which ray) and next-rank kernel (where a ray goes next) work exactly like in \brix, using \emph{proxies} built over data that is pre-partitioned across ranks 
using---in \vopat's case---some k-d tree partitioning. \vopat initially supported only structured data, but was later also extended to handle unstructured-mesh data. Rendering uses Woodcock tracking and can thus also support volumetric shadows, HDRI env-map lighting, and even volumetric scattering. For more technical details, we refer to the \brix paper~\cite{brix}, and to the freely available source code on GitHub~\cite{vopat}.

\medskip

\vopat was originally written before \rafi existed and already had dedicated ray-forwarding code. Adopting \rafi was not about creating a new piece of software, but rather an exercise in porting the original code to learn just how much simpler that software could become.

Most of \rafi did not require any changes: data importers, the viewer, scene partitioning, proxy computation, OptiX kernels for tracing rays against proxies, and all the render kernels for DDA marching, delta tracking, and scattering/shading remained unchanged. The only change in the device code is that all reading and writing of ray queues is now done through the \rafi device interface, resulting in simpler, less code. On the host side, adopting \rafi allowed for removing large amounts of MPI code and CUDA kernels---all replaced with a single \code{rafi->forwardRays()}. In summary, the ported \vopat uses significantly less (and more readable) code that would, in retrospect, have been much easier to write than the original.

A much more interesting question arises once we look at communication \emph{other} than forwarding rays. \vopat---like \brix---also uses a \emph{Distributed Frame Buffer} for gathering and distributed merging of different ranks' local frame buffers (also see~\cite{brix}). If \rafi had already been available when \vopat was written, this would probably have been solved by forwarding pixel contributions via \rafi, but deciding whether that is beneficial is less simple. On the one hand, handling frame buffer updates through \rafi would have been much easier to realize (which is good), but it may well have resulted in higher bandwidth and lower performance (which is not).

\begin{figure}[bt!]
\begin{center}
  \includegraphics[width=.98\columnwidth]{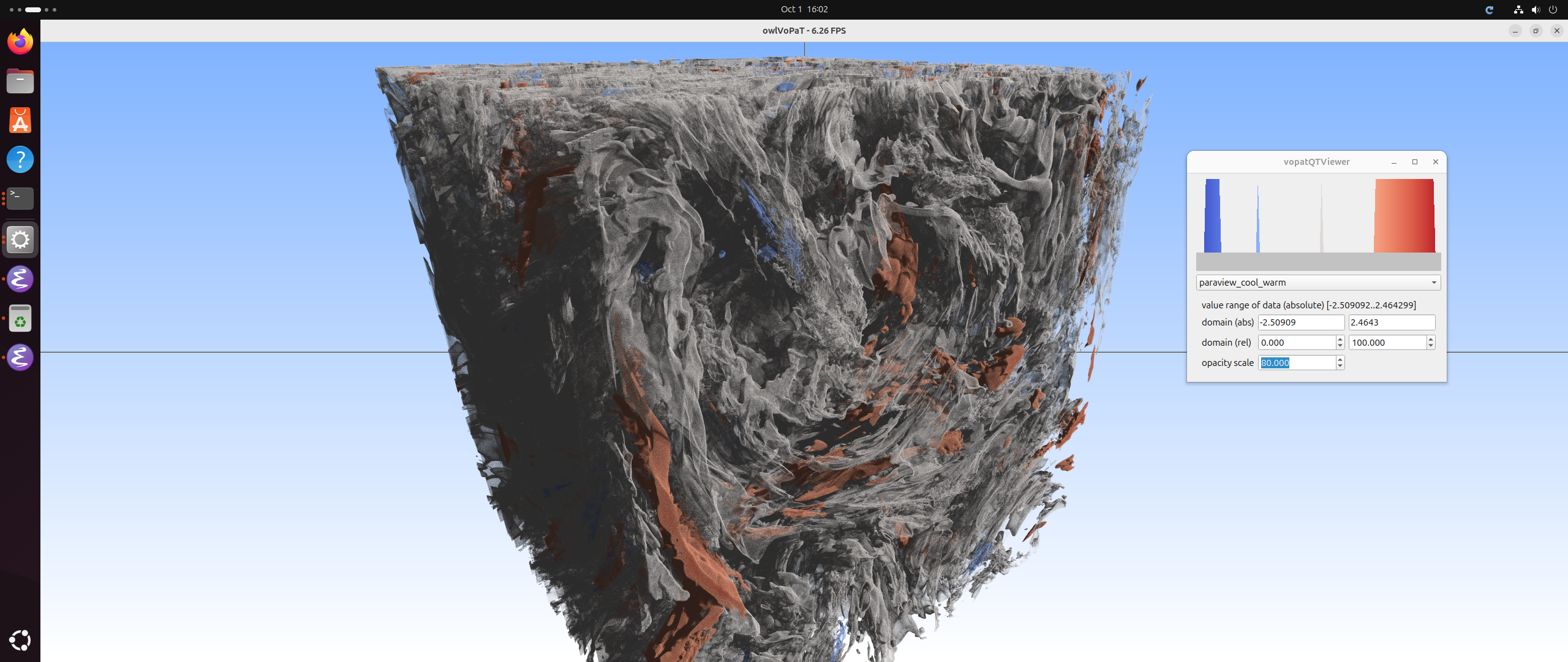}
  \\
  \includegraphics[width=.98\columnwidth]{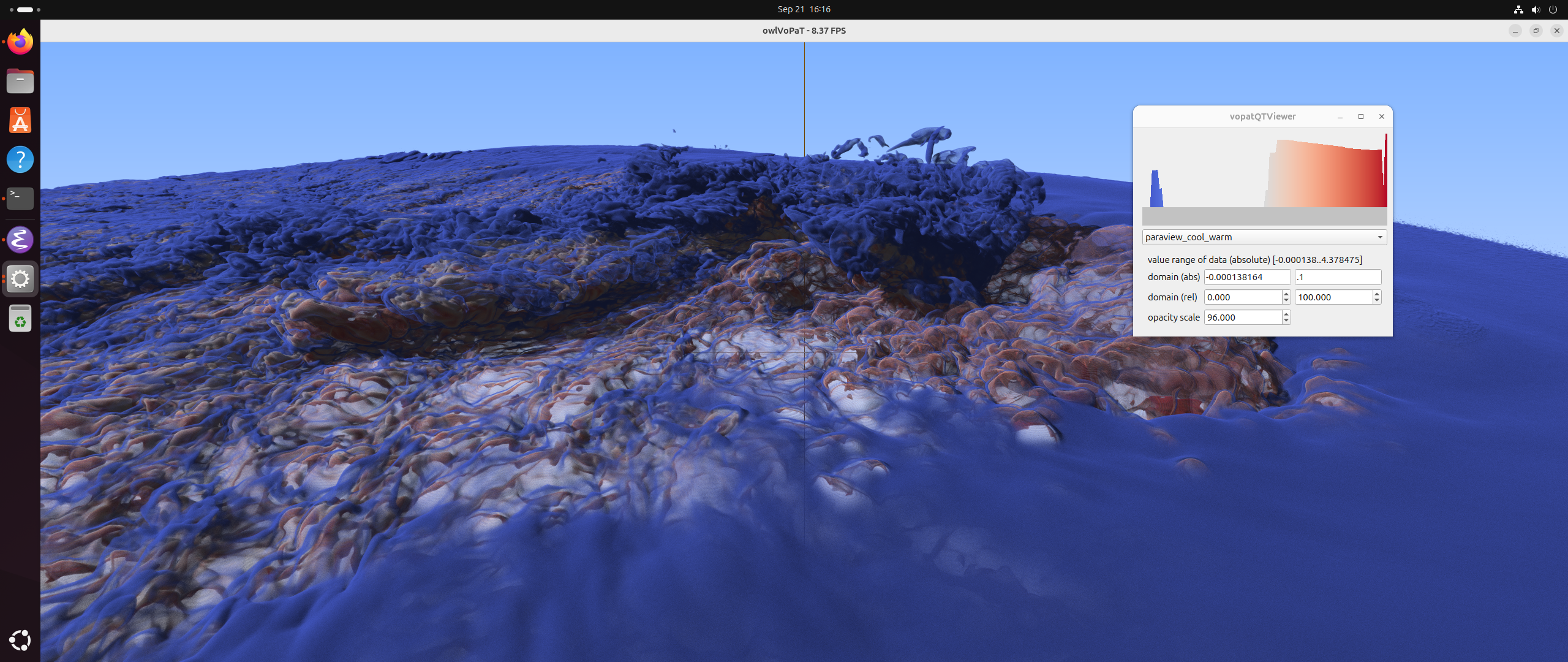}
  \\
\end{center}
\vspace*{-1.5em}
\caption{\label{fig:vopat} Two screenshots from our \rafi-enabled \vopat volume path tracer, running on 
four 10GB-Ethernet networked PCs with two RTX8000 (Turing) GPUs each.
Left: the $4K\times 4K\times 4K$ \code{rotstrat} data set. Right: the $305\times 3001\times 2501$ \code{thunderstorm} data set.}
\vspace*{-1em}
\end{figure}

\medskip

Some screenshots of our \rafi-enabled \vopat can be seen in Figure~\ref{fig:vopat}. Since \rafi does not in any way change which rays are traced---or what these rays do---the rendered images will not differ in any way from those rendered without \rafi; \rafi just makes for a significantly simpler realization of \vopat, nothing else. We
never performed a detailed before/after performance evaluation, but would not expect any major difference either way.

Overall, we conclude that had \rafi been available when \vopat was being developed, it would have significantly reduced the effort to build this system---though by how much is impossible to measure.

\subsection{Data-Parallel Unstructured Volume Rendering}
\label{sec:lander}

\begin{figure}[ht!]
\centering
\setlength{\tabcolsep}{0pt}
\begin{tabular}{c}
\hline
\hline 
Before: compositing fragments via ``deep compositing'' as used in~\cite{lander-distributed} \\
\hline
\hline \\ [-0.8em]
\includegraphics[width=.98\columnwidth,trim=4pct 20pt 8pct 0pct,clip]{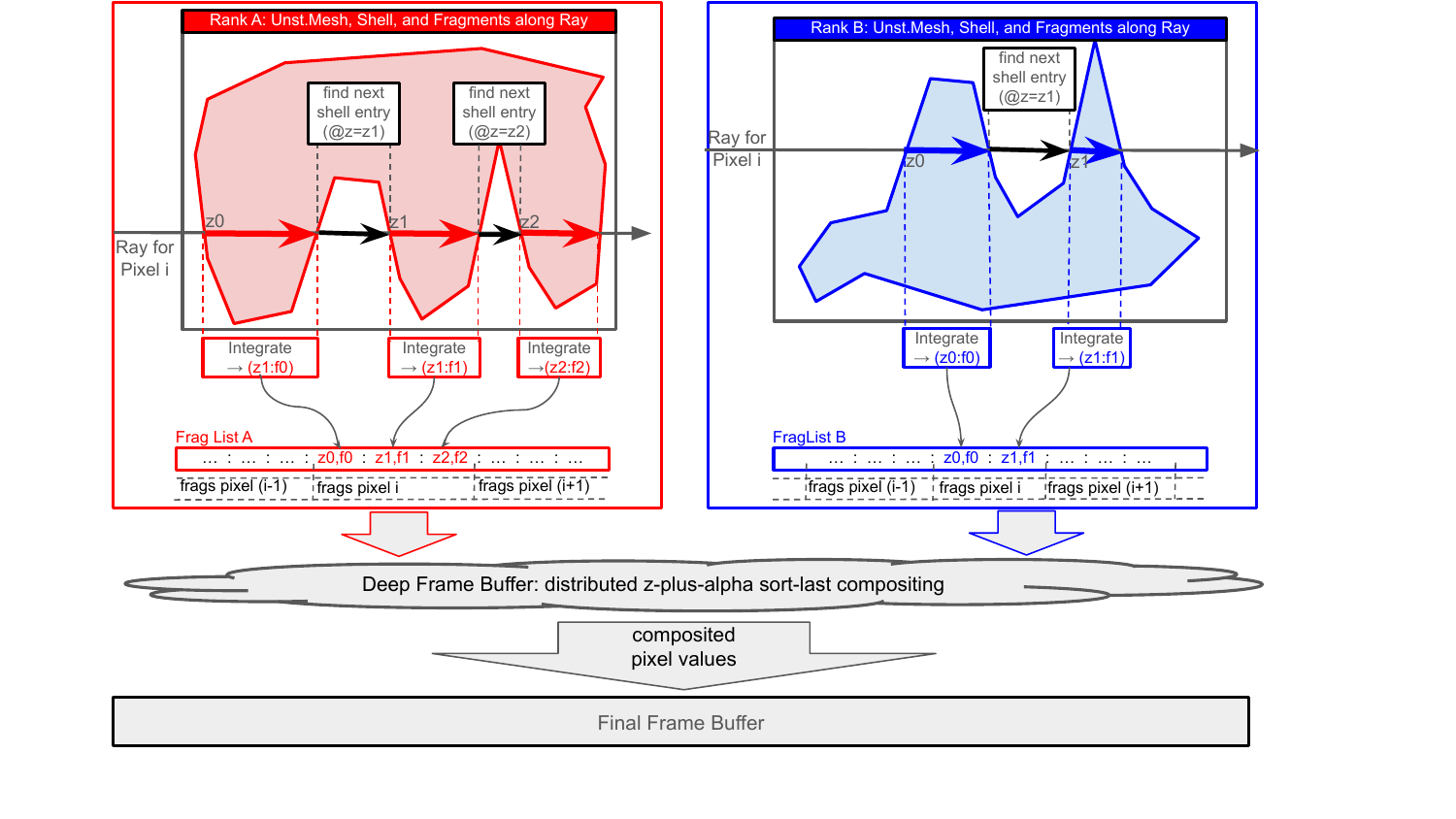}
  \\
  \hline
\hline
After: \rafi realization of the same problem \\
\hline
\hline
  \includegraphics[width=.98\columnwidth,trim=1pct 0pct 6pct 0pct,clip]{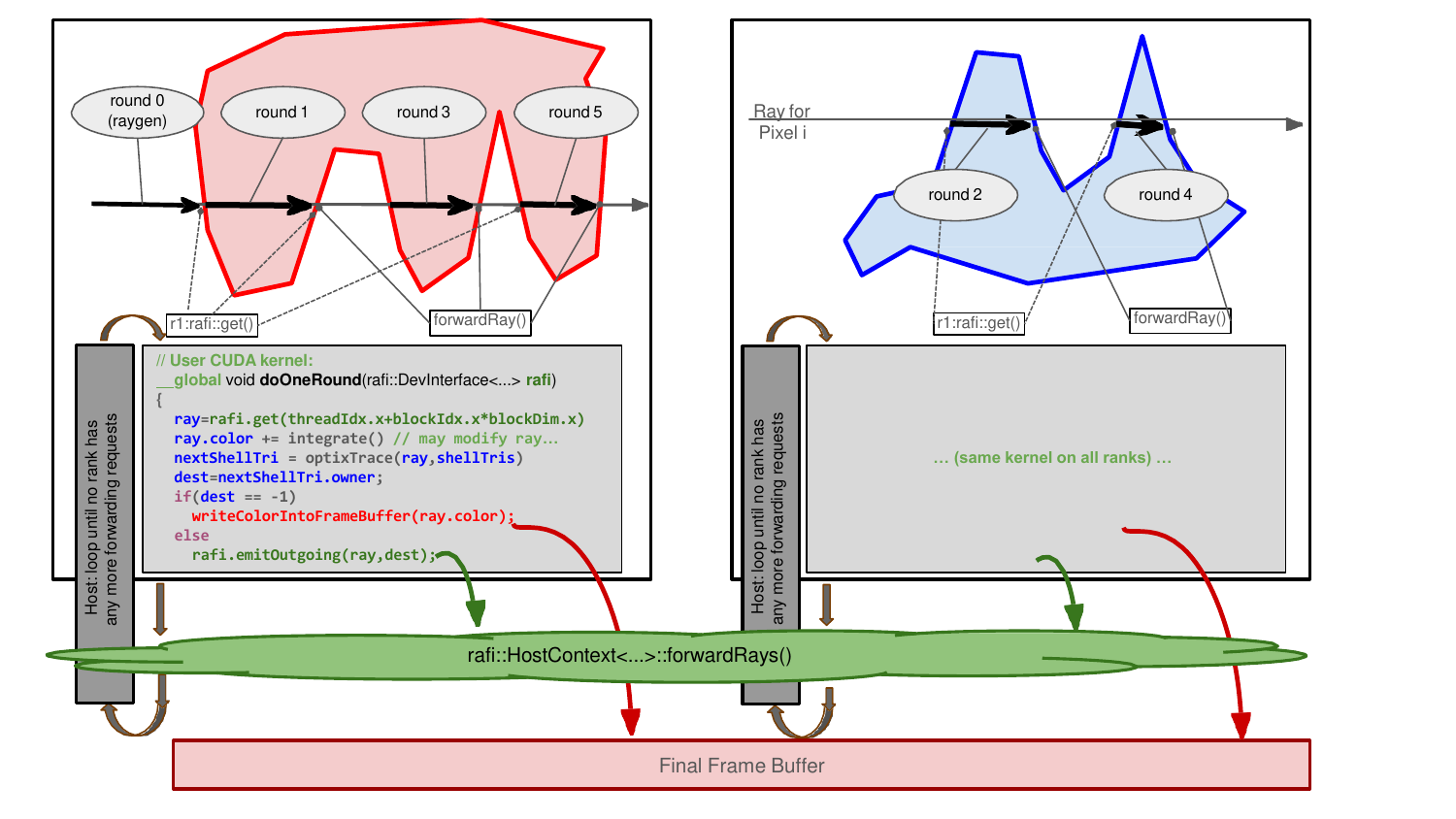}
\end{tabular}
    
\vspace*{-1.5em}
\caption{Illustration of before (top) vs.\ after adopting \rafi (bottom) for the in-situ Mars Lander rendering framework discussed in Section~\ref{sec:lander}. Both variants use the same data partitioning and the same shell traversal~\cite{sahistan2021Shell} to find where rays enter and leave a given rank's unstructured mesh. The original method (top) can only use straight rays and relies on "deep compositing" to alpha-composite the fragments created by different ray segments. The \rafi variant (bottom) instead forwards rays from rank to rank; this is easier, not limited by the number of segments, and can also---though not shown in this illustration---also change ray directions and spawn shadow rays (neither of which is allowed in compositing). 
\label{fig:lander-illustration} 
\vspace*{-1em}
}
\end{figure}

Sahistan et al.~\cite{lander-distributed} described a framework for the data-parallel volume rendering of natively distributed unstructured data. The goal of this framework was to eventually enable in-situ rendering of data within CFD codes such as LAVA or FUN3D; the main challenge it addressed (other than the sheer size and complexity of the data itself) was that the native data partitioning produced by such codes is often non-convex.

The problem with non-convex partitions is that any ray can enter and leave a given rank's partition multiple times. When volume rendering, each such entry-exit ray segment produces a separate color-plus-alpha \emph{fragment}. Traditional sort-last rendering can handle only a single fragment per pixel and thus cannot handle this situation. To handle this, Sahistan et al.\ proposed \emph{deep compositing}, which can handle more than one fragment per pixel. This method significantly improved upon traditional compositing but still had two limitations: one is that deep compositing could store only a fixed number of fragments per pixel and had to drop or out-of-order-merge fragments if that number was exceeded; this could lead to rendering artifacts for rays that straddled partition boundaries. A second limitation was that, like any other sort-last rendering technique, it only worked for rays that passed straight through the model, and thus could not handle either volumetric scattering or volumetric shadow.

To address these limitations, we ported this framework
to \rafi.
Primarily, \rafi interfaced with the boundary traversal logic---i.e., shell-to-shell traversal---of this application. On each rank, rays are dequeued from \rafi and traced through the local mesh using \optix over a face-based acceleration structure. The traversal advances a ray from one cell to the next by intersecting mesh faces. Upon encountering a partition boundary, the ray determines the neighboring rank that owns the adjacent cell, packages its current integration state, and forwards itself via \rafi. Rays reaching a model boundary are finalized and contribute to their corresponding pixel.

Before \rafi, the application instead identified entry and exit points per rank, integrated the volume segment locally, and stored the resulting contribution and depth for later compositing. With \rafi, this becomes an explicit ray-forwarding pipeline that eliminates sort-last compositing: each ray carries its accumulated color and opacity as it progresses across consecutive partitions. This simplifies the rendering model, enables non-straight rays, and removes the fragment management required by compositing. Figure~\ref{fig:lander-illustration} illustrates the integration.

\begin{figure}[ht]
    \centering
    \includegraphics[
        width=0.98\linewidth,
        trim=96pt 54pt 94pt 54pt,
        clip
    ]{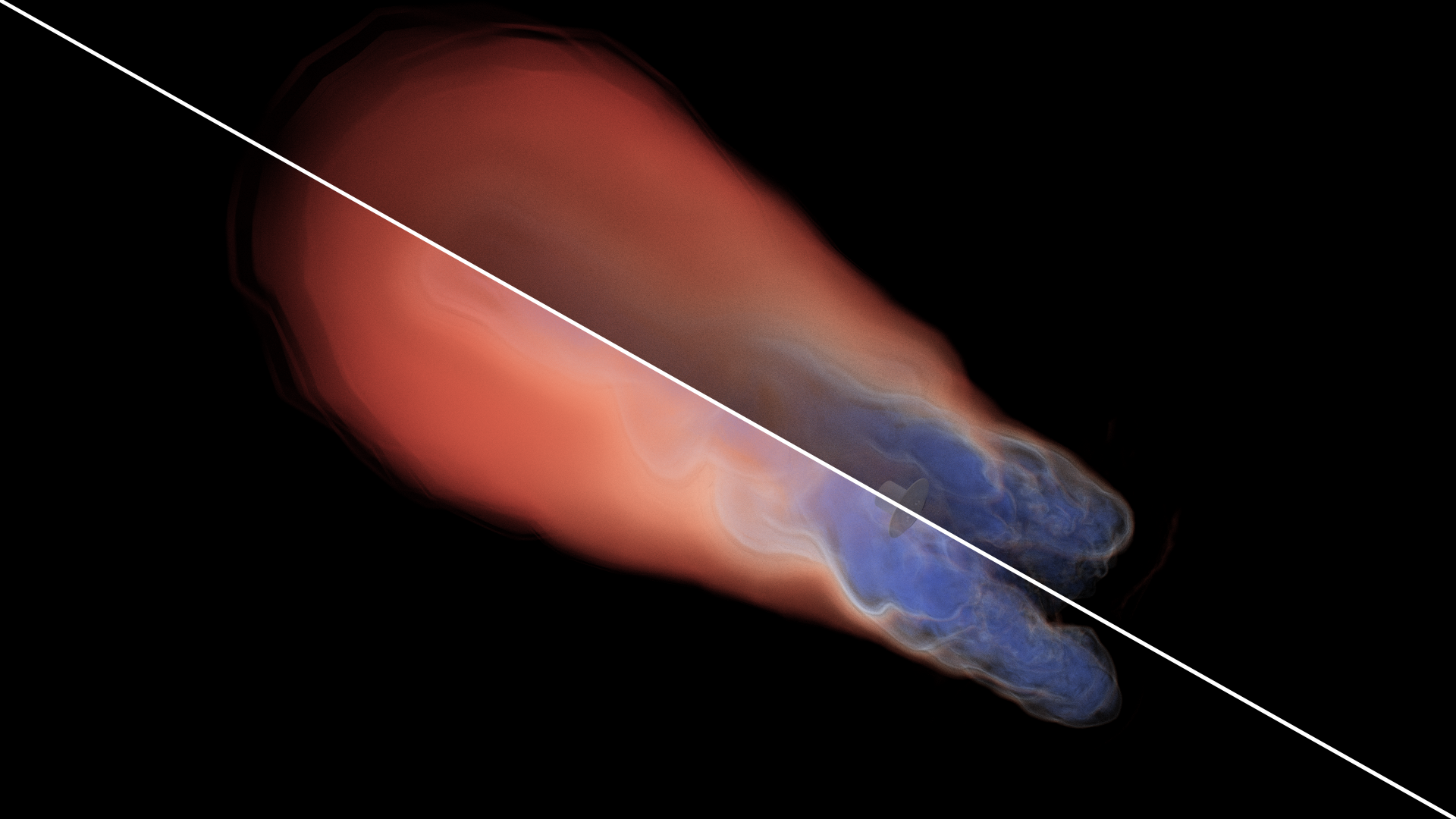}
        
\caption{Images of the 788 million unstructured
   element Mars Lander Retropulsion data set rendered with the \texttt{rafi/Lander}\xspace
   volume renderer from Section~\ref{sec:lander}.
   This uses 
    the native 
    native FUN3D data partitioning into 72 (highly non-convex)
    partitions, distributed across 16 MPI ranks (GPUs). Left: emission-absorption rendering (6.9 frames per second) that could also have been produced with a compositing-based renderer.
    Right: with volumetric shadows (6.1 frames per second), which compositing could not easily handle, but \rafi can.
    \label{fig:lander-shadow-on-off}
    }
\end{figure}

Although the implementation we discussed is based on~\cite{lander-distributed}, the volume integration uses delta tracking~\cite{fong:2017, Morrical2022QuickClusters}. Since \rafi supports non-straight rays, it also enables secondary rays—such as shadow rays—that were not possible with ``deep compositing'' (see Figure~\ref{fig:lander-shadow-on-off}).

The same shell-to-shell traversal and ray-forwarding infrastructure are reused in the Schlieren application described in Section~\ref{sec:schlieren}, with only the per-ray integration model changed.

\subsection{Data-Parallel Unstructured Schlieren Rendering}
\label{sec:schlieren}

In the previous section, we saw how \rafi supports the development of a volume rendering algorithm that could operate within a CFD solver as an {\em in situ} capability.  As an initial test case, we implemented support for distributed unstructured grids in the native domain decomposition used by FUN3D. In this section, we revisit the same type of distributed unstructured grid, but for a different type of rendering. {\em Schlieren} is a technique for visualizing inhomogeneities in transparent media such as air, water, or glass \cite{Settles:2001}. Toepler~\cite{toepler1864beobachtungen} first coined the term Schlieren and developed the method in the mid-19th century. In the latter 20th century, Schlieren became widely used for visualizing shocks in high-speed flows around aircraft and space vehicles. Yates~\cite{Yates:1993} was among the first to propose a technique for computing Schlieren images from CFD solution data. He argued that one can produce relatively accurate Schlieren images by sampling a domain on straight rays rather than tracing refracted rays. Most implementations described in the literature since Yates' seminal work have also been based on straight-ray algorithms. 

\begin{figure}[htbp]
    \centering
    \hfill
    \includegraphics[width=0.49\columnwidth]{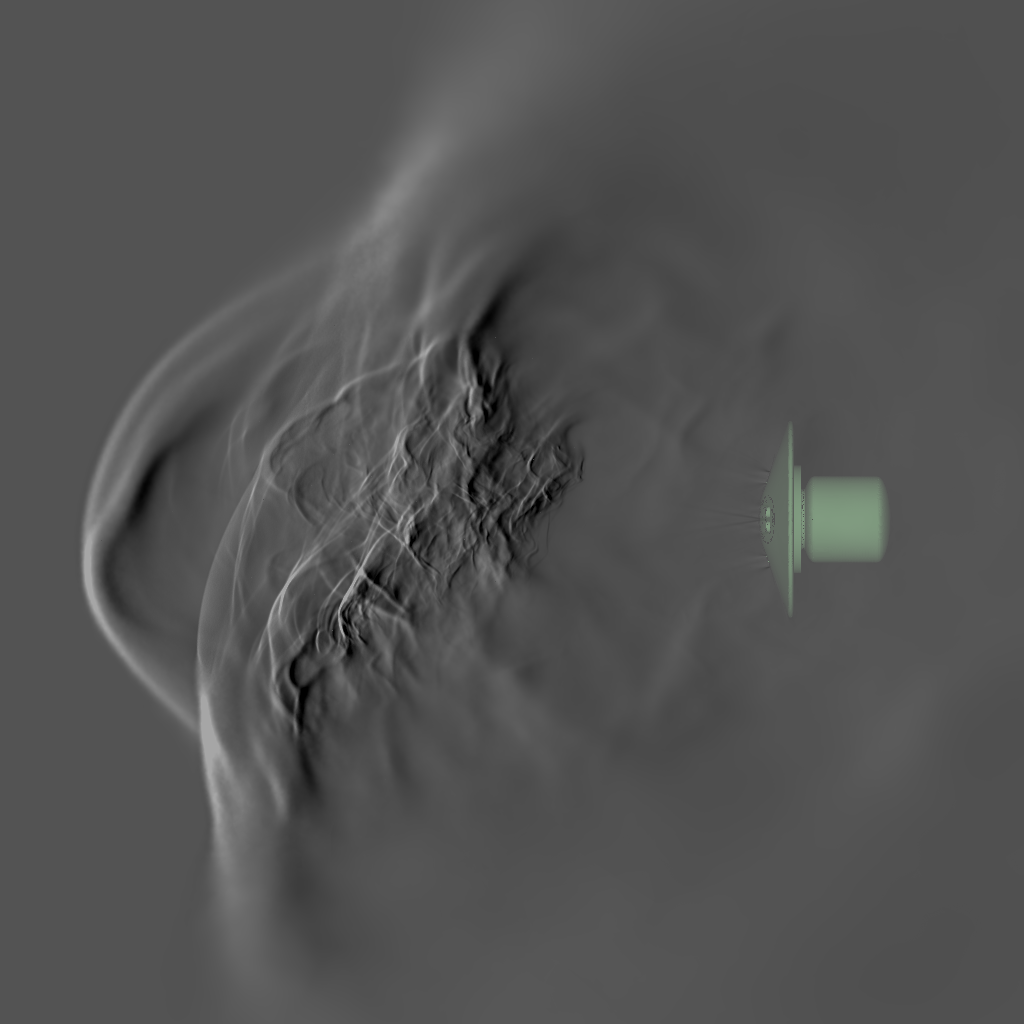}
    \hfill
    \includegraphics[width=0.49\columnwidth]{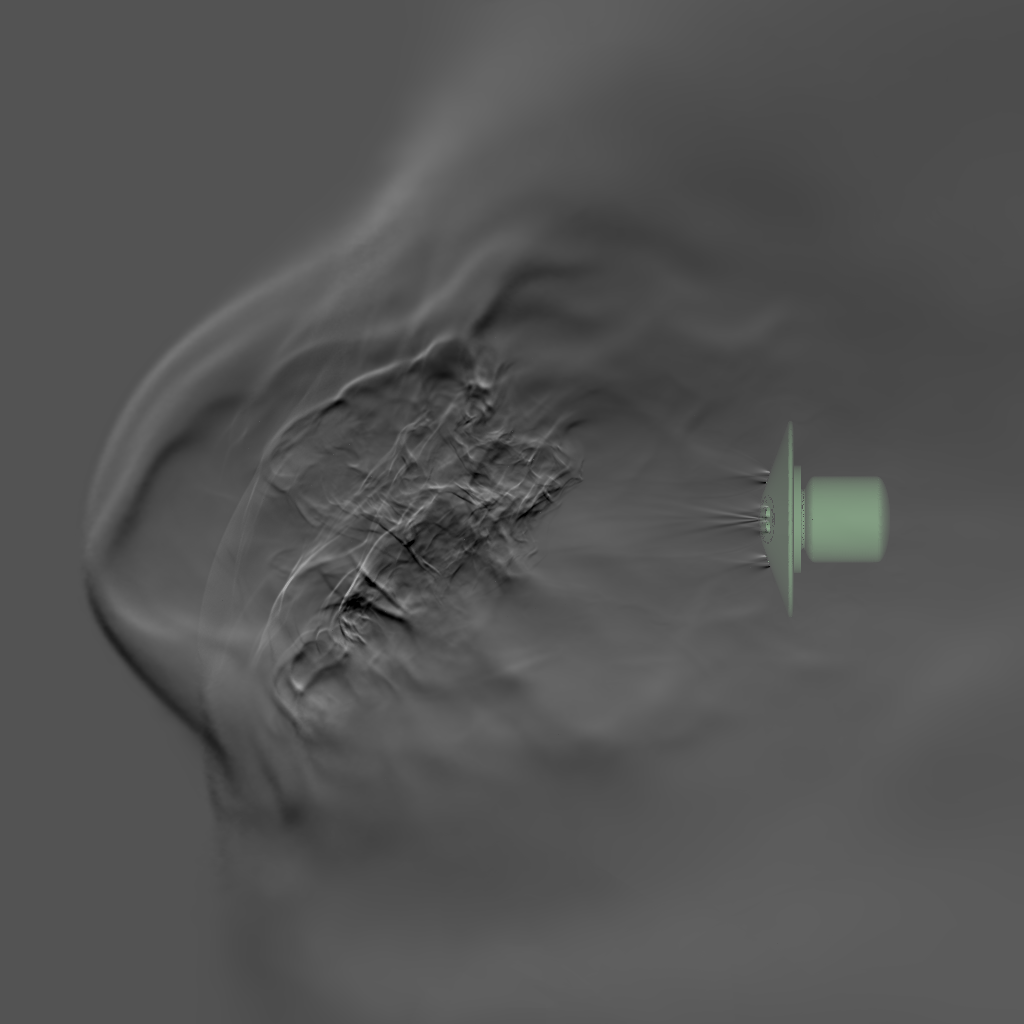}
    \hfill
    \\[-1ex]
    \caption{Two images from the \rafi-enabled \schlieren renderer described in Section~\ref{sec:schlieren}, in both cases showing the Mars Lander Retropropulsion data set. Left: knife-edge filter in the $u$ direction (emphasizing horizontal gradients with respect to the viewer). Right: knife-edge filter in the $v$ direction (emphasizing vertical gradients).}
    \label{fig:schlieren}
\end{figure}

As an initial effort, we ported a previously-developed
distributed-memory Schlieren
implementation based on a framework called {\em Slurry} to \rafi.
Incorporating \rafi allowed us to restructure this process into explicit ray forwarding while leaving the Schlieren integration, \optix traversal, and ``shading'' logic largely unchanged.  
At each rank, like slurry, we use OWL~\cite{owl} to build an \optix acceleration structure over the faces of an unstructured mesh that represents the underlying scalar field, including per-vertex density and density gradient attributes.  For faces on the outside of the mesh, we also store which rank owns the cell on the exterior side of this face (or -1 if it is on the model boundary).

Ray traversal follows the same shell-to-shell logic described in Section~\ref{sec:lander}. Rays are dequeued from \rafi, traced through local mesh faces using \optix, and either forwarded to the neighboring rank when hitting a partition boundary, or finalized at the model boundary. Each forwarded ray packages its entire persistent integration state (Listing~\ref{lst:schlieRafiDataStructs}).

\begin{lstlisting}[%style=cppstyle,
    float,
    floatplacement=htbp,
    caption={Forwardable ray state that \rafiSchlieren uses to describe what gets forwarded between partitions. },
                   label={lst:schlieRafiDataStructs}]
struct FWDRay {
//  --- Ray state ---
  vec3f origin;       // current position
  vec3f direction;    // ray direction
  float tmin;         // restart parameter
  int   pixelID;      // framebuffer index
// --- Partial results ---
  float integral;     // accum'ed Schlieren integral
  vec3f surfColor;    // surface contribution
};
\end{lstlisting}

After the \optix launch terminates, each rank calls \code{rafi->forwardRays()}, the result of which provides the input for the next \optix launch.

This process repeats until no rays remain across all ranks, at which point distributed termination is reached.

Figure~\ref{fig:schlieren} illustrates example results from our \rafi Schlieren renderer.  In optical Schlieren, a knife-edge is used to filter the light reaching a display screen. The knife's orientation controls the direction of emphasis.  Traditionally, a ``U'' image emphasizes horizontal gradients with respect to the viewer, and a ``V'' image emphasizes vertical gradients. We can recreate these knife-edge effects in the algorithm.

What \rafi buys us, compared to slurry, is two-fold: One is
the ability to use a more general library rather than a
special-purpose implementation, with its associated maintenance cost. Second, the \rafi version also
offers the generality needed to implement Schlieren algorithms that require tracing refracted rays. 
Though we have not yet used that
capability, the question of how much of a difference a refracting-ray Schlieren algorithm will make compared to a straight-ray approximation is one we have long been interested in investigating further.  With \rafi, we will be one step closer, particularly for large datasets that require a distributed-memory approach for processing.

\subsection{MPI Particle Tracing for Streamline Computation}
\label{sec:particle-tracking}

So far, we have focused on use cases for which \rafi was originally intended for: renderers exchanging rays via forwarding. Similar communication patterns, namely multiple GPUs potentially distributed across a network exchanging their workload between computational rounds, occur in many applications across physics and other computational sciences. Here, the GPUs perform a defined, usually homogeneous set of computations and then synchronously exchange their working sets and continue, possibly in rounds, until all work is done.

\begin{figure*}
    \centering
    \resizebox{1.00\textwidth}{!}{
    \includegraphics[height=.29\linewidth
    ]{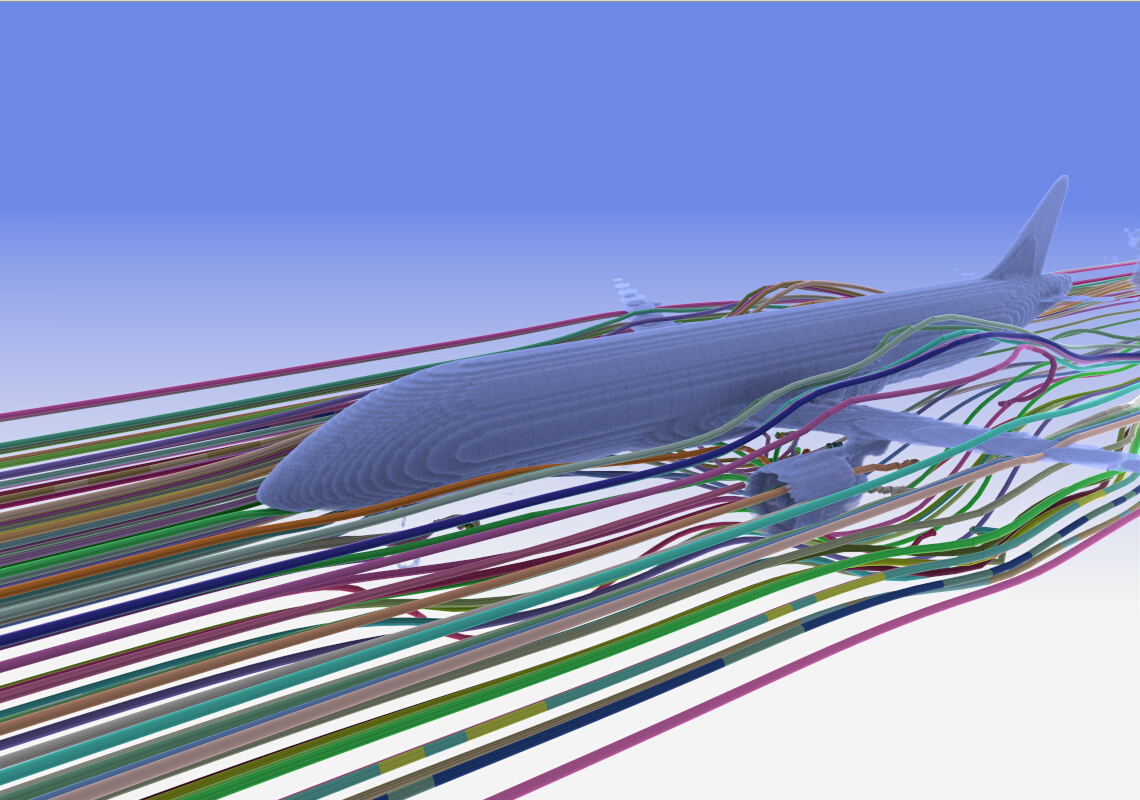}

    \includegraphics[height=.29\linewidth,trim=10pct 0pct 10pct 0pct,clip]{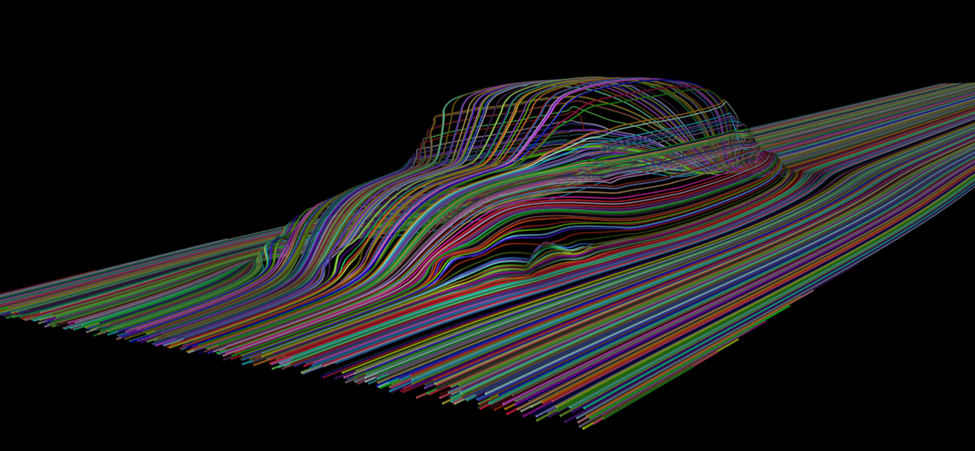}
    \includegraphics[height=.29\linewidth,trim=10pct 0pct 10pct 0pct,clip]{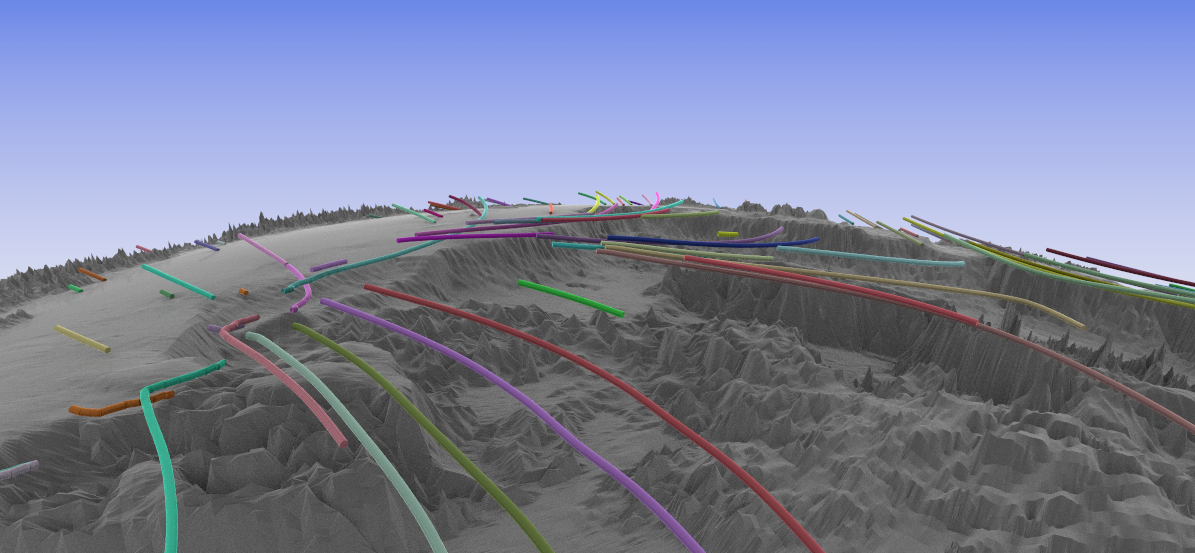}
    }\\[-1ex]
     \caption{\label{fig:streaklines}
    Streamlines computed with our \rafi-enabled
    data-parallel particle advection application described in
    Section~\ref{sec:particle-tracking}.
    Left: velocity field around an airplane computed with OpenFoam.
    Center: airflow around a car.
    Right: wind turbulence at low altitude from a daily snapshot
    computed with ICON, by the German weather forecasting institute (DWD).}
    \vspace*{-1em}
\end{figure*}

In interactive postprocessing, an algorithm with communication patterns similar to those presented before is particle tracing to compute streamlines (or streaklines and pathlines) using numerical integration. Here, the data is the output of a simulation and represents a 3D vector field sampled at discrete positions (e.g., on the corners of a uniform grid or unstructured mesh). In a data-parallel setting, given a spatial distribution (e.g., the one originally coming from the simulation), each rank is responsible for a portion of that distribution (e.g., a single macrocell in a grid). Given this a priori assignment of simulation output data to compute nodes, particles are advected and their paths are tracked. The particles form the working set, just as the rays did in previous applications. To visualize them, the particles on the same path are connected by line segments and rendered as cylinder or tube primitives (see Figure~\ref{fig:streaklines}).

The computation is round-based: in each round, the compute nodes are
assigned those of the $n$ particles advected that overlap with their
spatial domain; the number of rounds is either predefined or
terminates when no particles remain to be processed. Oftentimes, $n$
is comparably low to avoid visual clutter. This leads to load
imbalances, as at any given time, compute nodes often have no
particles to advect.

Given this setup and data distribution, in round~1, $n$ particles are released at predefined seed points so that each rank is responsible and keeps track of $0 \le k \le n$ particles at a time. Each rank/GPU independently performs an update step on each particle (i.e., one GPU thread per particle on this rank) using the Runge-Kutta algorithm, resulting in an updated particle position, or the computation is terminated if the particle did not move.

After each round, some particles on each rank got terminated, while others changed position; of the latter, yet another subset left the integration domain for which the rank is responsible. In that case, the particle either leaves the dataset's integration domain (resulting in termination) or its position overlaps with the domain of another rank. In that case, the application must track where the particles moved after each round, and use interprocess communication to send the particles to their new destinations while simultaneously receiving particles that moved into the domain of this rank.

This communication pattern---that we prototypically implemented in an application that can support multiple 3D field types---turns out to be very similar to the patterns used in ray tracers, only that the ``ray type'' in our application now becomes a single particle. That particle consists of a unique ID (so we can track and identify them across ranks), and a 3D position (\code{float3}). Runge-Kutta updates are computed in a CUDA kernel with one thread per particle, as though the application ran on a single GPU; only at the end of each update step, the CUDA kernel determines based on the new particle position its destination in the
network (which can be another node or GPU, or this GPU as the particle did not move across boundaries). How the kernel does that is an implementation detail; e.g., if the space partitioning uses a grid, the neighboring rank is found by projecting the position onto the grid, yielding a new macrocell ID that maps to an MPI rank. More advanced partitioning is possible, but this is entirely the application's responsibility. Given the particle \code{P}, and the destination, the CUDA kernel calls \code{rafi.emitOutgoing(P, destination)} and returns; after the round finishes, the host calls \code{rafi->forwardRays()} to exchange its ``rays'' (or rather: particles) and initiates the next round.

\subsection{Distributed Barnes-Hut N-Body Computation}
\label{sec:nbody}

As a final test case, we also used \rafi to implement a distributed $N$-body simulation running on multiple GPU ranks with CUDA-aware MPI. Distributed N-body is a non-trivial challenge from a communication perspective: not only must particles migrate between ranks as they cross domain boundaries, but each rank also needs multipole information from remote domains before it can compute forces. Without a framework like \rafi, realizing this requires hand-written MPI communication for at least three distinct data types and communication patterns: particle migration, essential-tree exchange, and refinement requests, each with its own serialization, routing, and synchronization logic. Our implementation employs the Barnes-Hut (BH) tree~\cite{barnes1986hierarchical} with quadrupole corrections. All inter-rank communication is handled exclusively through \rafi contexts, making the host code straightforward despite the multi-phase protocol. 

The domain is partitioned using a Morton-order decomposition, which has the useful property that the owner rank of any particle position can be computed directly on the GPU with no CPU-side routing tables.
On each rank, a \cuBQL BVH~\cite{cubql} is built over the locally owned particles, and a bottom-up pass then aggregates per-node center-of-mass and quadrupole data for use in the Barnes-Hut multipole approximation.
Time integration uses a leapfrog scheme.

\begin{lstlisting}[caption={The three \rafi \code{Ray} types in the N-body application.},
label={lst:nBodyRafeTypes}]
// Used in the Particle exchange stage
struct Particle {
  float3 pos;        // position
  float3 vel;        // velocity
  float3 force;      
  float  mass;
};
// Used in the Tree Exchange stage
struct VirtualParticle {
  float3 pos;        // center of mass
  float  mass;
  float  smax;       // node size for MAC; 0 = leaf
  int    sourceRank; // originating rank
};
// Used in the Tree Exchange stage
struct RefinementReq {
  int senderRank; // rank requesting refinement
};
\end{lstlisting}

An important aspect of this application is that it has different stages that need to communicate via different data. We do this by using three distinct \rafi instantiations, each with a different ``Ray'' type reflecting the fundamentally different pieces of information that travel between ranks (see Listing~\ref{lst:nBodyRafeTypes}). The \code{Particle} context handles migration: after each integration step, each GPU thread determines the next rank on-device and calls \code{rafi.emitOutgoing}, and the data that travels in this stage is \code{Particle}s. 
The \code{VirtualParticle} and \code{RefinementReq} contexts (Listing~\ref{lst:nBodyRafeTypes}) together drive an adaptive essential-tree exchange: each rank first broadcasts its root multipole node to all peers. Peers that need finer remote data send back a \code{RefinementReq}, and the responding rank emits the relevant subtree nodes as additional \code{VirtualParticle}s. The combined remote multipole data is then used to build a second BVH for force computation. In this stage, the data communicated are \code{VirtualParticle}s and \code{RefinementReq}s. This example demonstrates that \rafi generalizes well beyond ray tracing and, through its support for multiple simultaneous instantiations with different data types, can cleanly express multi-phase distributed algorithms.

\begin{figure}
    \centering
    \includegraphics[width=.48\linewidth]{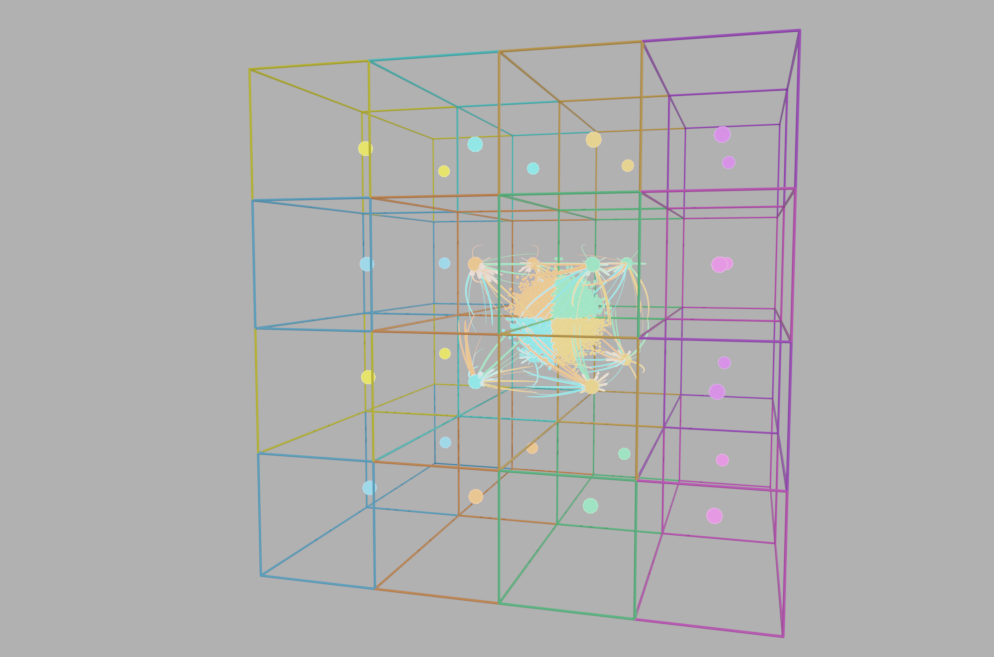}
    \includegraphics[width=.48\linewidth]{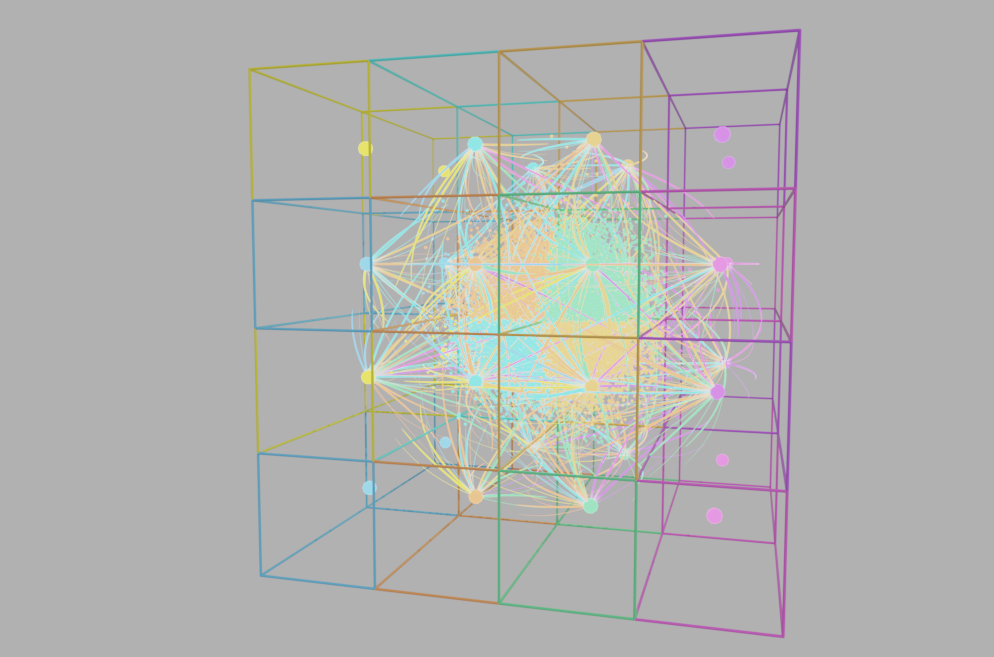}\
    \\[-1ex]
        \caption{Visualization of the communication patterns
        in our \rafi-enabled N-Body simulation from Section~\ref{sec:nbody}, for
        two different simulation steps. Boxes correspond to the different ranks'
        parts of the domain; connections show which ranks exchange messages.}
        \vspace*{-1em}
    \label{fig:nbody_mpi}
\end{figure}

\section{Summary and Discussion}

We have described \rafi, a software infrastructure that enables the easy forwarding of rays---or similar items of work---between GPUs in various data-parallel render or compute kernels. The defining feature of \rafi is how much easier it makes data-parallel computation, where ranks must semi-randomly forward work items between GPU kernels running on different ranks. We have shown this in multiple examples: For \vopat (Section~\ref{sec:vopat}), the ported application does exactly the same as the original, but with significantly
less and easier code. Had \rafi already existed when \vopat was first built, doing so would have been much easier. 

For \raldr (Section~\ref{sec:lander}), the original application (i.e., before porting to \rafi) 
had to rely on sort-last compositing using a deep frame buffer---which sometimes caused rendering artifacts along rank boundaries (when the number of fragments exceeded the depth of the deep frame buffer), and
which was intrinsically limited to simple emission-absorption rendering without shadows or scattering. 
Porting to \rafi allowed for eliminating both of these shortcomings, allowing for both volumetric shadows
and correct handling of rays iterating through any number of domain boundaries without artifacts. The
price for this is somewhat higher communication cost in both data volume and latency. While we have not
seen a huge performance impact in our particular application, it is nevertheless worthwhile to point out
that improving capabilities isn't always free.

For \rafiSchlieren (Section~\ref{sec:schlieren}), the original application also relied on compositing, and since Schlieren imaging---unlike volume rendering---works by \emph{adding} rather than compositing different ranks' images, this did not require a deep frame buffer, nor did it come with any limitations. Porting to \rafi will eventually allow experimenting with non-straight rays. Since virtually all state-of-the-art applications use straight rays, it is not yet clear how impactful this will be, but even
answering that question will be useful. 

Finally, in \rafiStreamLines and \rafiNBody we primarily showed how easy it was to build new data-parallel applications that didn't exist before. Obviously, either one of those could also have been written without \rafi, by simply writing the appropriate MPI code---but doing so is not always trivial, and with \rafi, it was.

\subsection{Performance}

So far, we have described the software architecture of \rafi,  how different examples use it, and how adopting \rafi made these applications simpler, easier, or more capable. However, performance assessment depends on various parameters. 

\begin{figure}
    \centering
    \includegraphics[width=\linewidth]{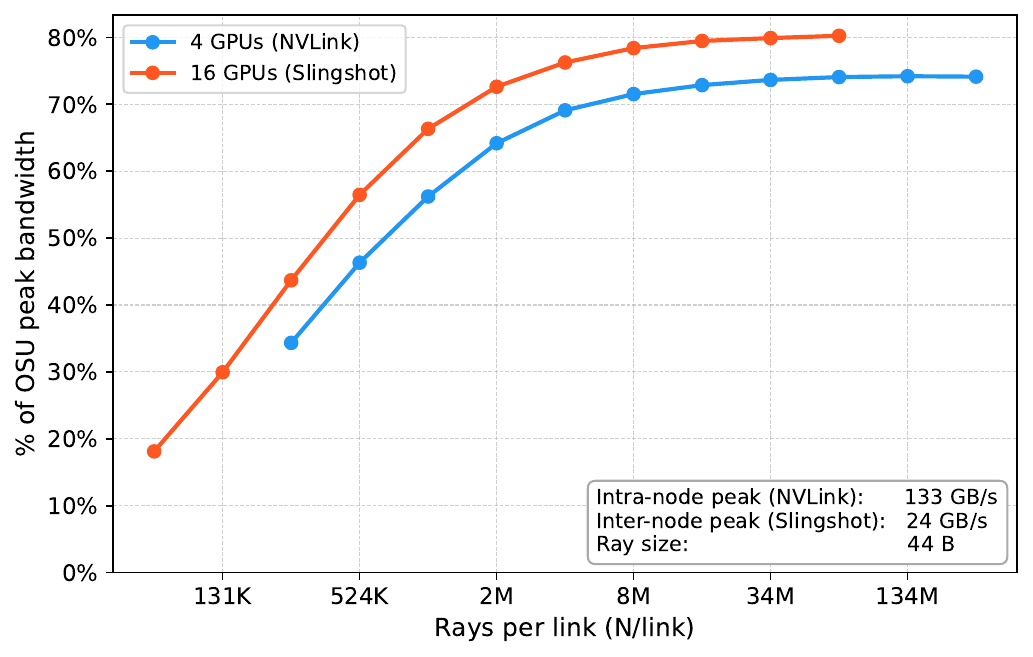}
    \\[-1em]
    \caption{Communication bandwidth utilization as a function of \texttt{Ray} count per link for Rafi-based exchanges on Alps. The intra-node configuration (4 GPUs per node, NVLink) and inter-node configuration (16 GPUs, HPE Slingshot) are shown as percentages of their respective OSU point-to-point benchmarks. Each ray payload is 44 bytes. Bandwidth efficiency increases with problem size in both regimes, reaching approximately 75\% of inter-node peak and 80\% of intra-node peak.\vspace*{-1em}}
    \label{fig:bandwidth}
\end{figure}

What is potentially interesting to the user is the effective forwarding bandwidth and, based on that, the overhead that using \rafi incurs to the app. To that end, we performed a benchmark for ray forwarding only, on the Alps supercomputer at CSCS, using NVIDIA GH200 Grace Hopper nodes interconnected via HPE Slingshot (cf.~Figure~\ref{fig:bandwidth}). In an intra-node configuration, we measure a sustained sort-and-send throughput of approximately 100 GB/s, 75\% 
of the measured NVLink per-device peak, corresponding to roughly 2.1 billion 44-byte rays per second. On an inter-node configuration, we sustain 20 GB/s, 80\%
of the 24 GB/s per device peak, or approximately 500 million rays per second (per link). In conclusion, we observe that the overhead in bandwidth only is negligible, as any parallel application would have to exchange its workload in one way or another.

By far more important, though, is how exactly the application uses \rafi, and the communication patterns, bandwidth, and other characteristics that this entails. \rafi might make it easier to experiment with different communication patterns, but in itself, it will not reduce any communication bottlenecks, bandwidth, or starvation issues.

On the upside, what we can say is that all else being equal, \rafi will carry little overhead of its own: On the device, the cost of writing into the \rafi queues is a single atomic-add, plus the cost of writing the work item, both of which are negligible on a modern GPU. Most of the actual work of \rafi happens in \code{rafi::forwardRays()}, and even that is mostly just calling CUDA kernels for generating the sort key, radix-sorting those int64 keys, and re-arranging the rays based on that key---and again, all of those are trivially cheap on a modern GPU. The total cost of what \rafi does is thus largely determined by the cost of sending the rays over the network---which the application would have to incur whether it uses \rafi or not. 

On the downside, the availability of an easy-to-use work-forwarding paradigm may encourage applications to adopt more costly communication patterns than they would otherwise. A case in point is our \schlieren example, where sending rays is absolutely going to be more expensive than each rank computing the integral of a Schlieren function on the rays that pass through it and then adding those results to get Schlieren function integrals over the full length of each ray, as we did with our slurry implementation. In this case, both the original (compositing-based) and the new (\rafi-enabled) Schlieren code currently produce the same answer, but the \rafi variant will absolutely carry a somewhat higher communication cost---not because \rafi is slow, but the most convenient way to automatically implement this code via \rafi incurs more communication than a more application-specifically crafted solution might have.

Yet another example of this is \vopat: had \rafi been available when it was first written, it would most likely have used \rafi for forwarding pixel fragments rather than having some dedicated distributed frame buffer---this would probably not have been catastrophic (the ray bandwidth would still have been larger), but it would nevertheless have produced higher bandwidth.

\subsection{Usage Beyond \emph{Ray} Forwarding}

As its name implies, \rafi was originally conceived in the context of \emph{ray} forwarding to simplify the creation of systems such as \brix, \vopat, and \barney. However, as we have shown above, it is not at all limited to rays
or rendering, and in general formulates the forwarding of any other sort of \emph{work}. Both the particle tracking (Section~\ref{sec:particle-tracking}) and nBody applications (Section~\ref{sec:nbody}) are already non-rendering workloads. Though we eventually also visualized their outcomes, the actual problems solved with \rafi were not rendering-related.

\subsection{Limitations}

One of \rafi's limitations is that it cannot relieve the user of the problem of \emph{capacity} management; i.e., \rafi manages the allocation of ray queues, but the user still needs to specify a maximum queue size.
For the applications we sketched above, it was always possible to compute an upper bound on the number of rays that could be active on any node at any point in time; so queues could be sized accordingly. How to do this depends on the application, however, so \rafi cannot relieve the user of this responsibility.

A similar restriction is that \rafi has the context of \emph{forwarding} work, and can, to some degree, also specify that a certain incoming work item can spawn more than one outgoing work item (e.g., a ray being shaded can spawn both a shadow ray and a bounced ray as well). However, it has no equivalent of MPI's \emph{reduce} operations. If some of the spawned items need any sort of reduction between themselves, \rafi cannot help with that; \rafi will only transport rays, and nothing more. Similarly, as discussed above, \rafi will only make it easier to develop the codes that need work forwarding; it does not inherently address issues such as bottlenecks, starvation, or long-tail problems that may arise in such implementations.

Finally, \rafi works best for a certain family of algorithms that rely on forwarding individual items of work between ranks in semi-random ways. For applications where the communication patterns are more clearly defined, it may be both easier and more efficient to use alternative communication methods. For example, while \rafi fits the ray forwarding patterns of the \brix renderer, the NVIDIA Barney renderer instead uses \emph{ray queue cycling}~\cite{RQS}, in which every rank always communicates with exactly one other rank. This could still be realize that through \rafi, but there is no real reason to do so.

\balance
\section{Conclusion and Future Work}

In this paper, we have discussed \rafi, a general ray (and similar work)-forwarding library for data-parallel GPU computing. While \rafi's original motivation was in ray-forwarding-based path tracing, it is not restricted to rendering, as we have shown in various examples. The main advantage of \rafi is its generality, and how much easier it makes it to write such work-forwarding applications. So far, we are very happy with how \rafi has worked out: Every one of our sample applications was created by a different set of developers, and not a single one encountered any issues worth mentioning. 

Across the topics covered in this paper, there is potential for interesting future work: For \schlierafi, the stated goal is to use our current prototype to develop an in-situ Schlieren imaging module for LAVA and/or FUN3D. This will require significantly more work concerning APIs and AMR data, but \rafi will likely remain a key piece in that effort. It would also be interesting to experiment with non-straight rays in Schlieren imaging. For our more volume-rendering-oriented examples, turning them into widely used libraries is less clear, as these would have to compete with existing data-parallel rendering libraries like Barney.

For \rafi itself, an interesting angle to explore is the use of other communication platforms such as NCCL. For truly high-performance applications, one would likely also want to consider extending \rafi to support multiple asynchronous streams, but this request has not yet come up. Lastly, it would be interesting to see if \rafi could
possibly be integrated into VTK---not only as an alternative to image compositing via Ice-T, but as a general data-parallel VTK communication primitive.

\acknowledgments{This work was funded by the German Federal Ministry of Research, Technology and Space under the funding code~01LK2204A. The responsibility for the content of this publication lies with the authors. This work was also supported by the Ministry of Education, Youth and Sports of the Czech Republic through the e-INFRA CZ (ID:90254). Additionally, this work was also supported by a grant from the Swiss National Supercomputing Centre (CSCS) for Alps.}

\bibliographystyle{abbrv-doi-hyperref}

\bibliography{main}

\begin{thebibliography}{10}

\bibitem{cudampi}
S.~Abbott.
\newblock {GPUDirect}, {CUDA-Aware} {MPI}, and {CUDA} {IPC}.
\newblock Oak Ridge Leadership Facility, Summit Training Workshop,
  \url{https://www.olcf.ornl.gov/wp-content/uploads/2018/12/summit\_workshop\_CUDA-Aware-MPI.pdf},
  2018.

\bibitem{abram2018galaxy}
G.~Abram, P.~Navrátil, P.~Grossett, D.~Rogers, and J.~Ahrens.
\newblock Galaxy: Asynchronous ray tracing for large high-fidelity
  visualization.
\newblock In {\em Proceedings of the IEEE 8th Symposium on Large Data Analysis
  and Visualization}, LDAV~'18, 2018.

\bibitem{HIP}
{Introduction} to the {HIP} programming model, 2026.
\newblock Available at \url{https://rocm.docs.amd.com/projects/HIP}, Last
  accessed: 29 March 2026.

\bibitem{barnes1986hierarchical}
J.~Barnes and P.~Hut.
\newblock A hierarchical {O (N log N)} force-calculation algorithm.
\newblock {\em Nature}, 324(6096), 1986.

\bibitem{slang}
N.~Bickford, T.~Lorach, and C.~Hebert.
\newblock Hands-on class: Introduction to slang: The next generation shading
  language.
\newblock In {\em Proceedings of the Special Interest Group on Computer
  Graphics and Interactive Techniques Conference Labs}, 2025.
\newblock SIGGRAPH Labs '25.

\bibitem{pprBook}
A.~Chalmers, E.~Reinhard, and T.~Davis.
\newblock {\em {Practical} {Parallel} {Rendering}}.
\newblock CRC Press, 2002.

\bibitem{chandy-lamport}
K.~M. Chandy and L.~Lamport.
\newblock {Distributed Snapshots: Determining Global States of Distributed
  Systems}.
\newblock {\em ACM Transactions on Computing Systems}, 3(1):63–75, Feb. 1985.
  \href{https://doi.org/10.1145/214451.214456}
{doi: {{%
10\hspace{.1pt}\discretionary{.}{%
}{.}\hspace{.4pt}1145\discretionary{/}{%
}{/}214451\hspace{.1pt}\discretionary{.}{%
}{.}\hspace{.4pt}214456}}}


\bibitem{childs2020terminsitu}
H.~Childs~et al.
\newblock {A} {Terminology} for {In} {Situ} {Visualization} and {Analysis}
  {Systems}.
\newblock {\em The International Journal of High Performance Computing
  Applications}, 34(6), 2020.

\bibitem{fong:2017}
J.~Fong, M.~Wrenninge, C.~Kulla, and R.~Habel.
\newblock {Production} {Volume} {Rendering}: {SIGGRAPH} 2017 {Course}.
\newblock In {\em ACM SIGGRAPH 2017 Courses}, SIGGRAPH '17, 2017.
  \href{https://doi.org/10.1145/3084873.3084907}
{doi: {{%
10\hspace{.1pt}\discretionary{.}{%
}{.}\hspace{.4pt}1145\discretionary{/}{%
}{/}3084873\hspace{.1pt}\discretionary{.}{%
}{.}\hspace{.4pt}3084907}}}


\bibitem{opencl}
J.~Kim, S.~Seo, J.~Lee, J.~Nah, G.~Jo, and J.~Lee.
\newblock {OpenCL} as a unified programming model for heterogeneous {CPU/GPU}
  clusters.
\newblock {\em SIGPLAN Not.}, 47(8), 2012.

\bibitem{cudaaware}
J.~Kraus.
\newblock {An} {Introduction} to {CUDA-Aware} {MPI}.
\newblock NVIDIA Technical Blog,
  \url{https://developer.nvidia.com/blog/introduction-cuda-aware-mpi/}, 2013.

\bibitem{cuda}
\mbox{NVIDIA Corporation}.
\newblock {CUDA Toolkit}.
\newblock Available at \url{https://developer.nvidia.com/cuda-toolkit}.

\bibitem{molnar1994sortingclass}
S.~Molnar, M.~Cox, D.~Ellsworth, and H.~Fuchs.
\newblock A sorting classification of parallel rendering.
\newblock {\em IEEE Computer Graphics and Applications}, 14(4):23--32, 1994.

\bibitem{icet}
K.~Moreland, W.~Kendall, T.~Peterka, and J.~Huang.
\newblock {An} {Image} {Compositing} {Solution} at {Scale}.
\newblock In {\em Proceedings of International Conference for High Performance
  Computing, Networking, Storage and Analysis}, 2011.

\bibitem{Morrical2022QuickClusters}
N.~Morrical, A.~Sahistan, U.~Güdükbay, I.~Wald, and V.~Pascucci.
\newblock {Quick} {Clusters}: {A} {GPU}-{Parallel} {Partitioning} for
  {Efficient} {Path} {Tracing} of {Unstructured} {Volumetric} {Grids}.
\newblock {\em IEEE Transactions on Visualization and Computer Graphics},
  29(01), 2023. \href{https://doi.org/10.1109/TVCG.2022.3209418}
{doi: {{%
10\hspace{.1pt}\discretionary{.}{%
}{.}\hspace{.4pt}1109\discretionary{/}{%
}{/}TVCG\hspace{.1pt}\discretionary{.}{%
}{.}\hspace{.4pt}2022\hspace{.1pt}\discretionary{.}{%
}{.}\hspace{.4pt}3209418}}}


\bibitem{navratil::PhD}
P.~A. Navratil.
\newblock {\em {Memory-Efficient, Scalable Ray Tracing}}.
\newblock PhD thesis, University of Texas, Austin, 2010.

\bibitem{cubql}
{cuBQL} - {The} {CUDA} {BVH} {Build} and {Query} {Library}.
\newblock Available at \url{https://github.com/NVIDIA/cuBQL}, Accessed: Feb 26,
  2026.

\bibitem{owl}
{OWL -- A Productivity Library for OptiX}, 2026.
\newblock Available at \url{https://github.com/NVIDIA/owl}, Accessed: Feb 26,
  2026.

\bibitem{park2018spray}
H.~Park, D.~Fussell, and P.~Navrátil.
\newblock {SpRay:} speculative ray scheduling for large data visualization.
\newblock In {\em Proceedings of IEEE 8th Symposium on Large Data Analysis and
  Visualization}, LDAV~08, pp. 77--86, 2018.

\bibitem{reinhard::PhD}
E.~Reinhard.
\newblock {\em {Scheduling and Data Management for Parallel Ray Tracing}}.
\newblock PhD thesis, University of East Anglia, 1995.

\bibitem{sahistan2021Shell}
A.~Sahistan, S.~Demirci, N.~Morrical, S.~Zellmann, A.~Aman, I.~Wald, and
  U.~G{\"u}d{\"u}kbay.
\newblock {Ray-traced Shell Traversal of Tetrahedral Meshes for Direct Volume
  Visualization}.
\newblock In {\em Proceedings of the IEEE Visualization Conference-Short
  Papers}, VIS~'21, 2021. \href{https://doi.org/10.1109/VIS49827.2021.9623298}
{doi: {{%
10\hspace{.1pt}\discretionary{.}{%
}{.}\hspace{.4pt}1109\discretionary{/}{%
}{/}VIS49827\hspace{.1pt}\discretionary{.}{%
}{.}\hspace{.4pt}2021\hspace{.1pt}\discretionary{.}{%
}{.}\hspace{.4pt}9623298}}}


\bibitem{lander-distributed}
A.~Sahistan, S.~Demirci, I.~Wald, S.~Zellmann, J.~Barbosa, N.~Morrical, and
  U.~G{\"u}d{\"u}kbay.
\newblock {Visualization} of {Large} {Non-Trivially} {Partitioned}
  {Unstructured} {Data} with {Native} {Distribution} on {High-Performance}
  {Computing} {Systems}.
\newblock {\em IEEE Transactions on Visualization and Computer Graphics},
  31(9):5000--5014, September 2025.

\bibitem{Settles:2001}
G.~S. Settles.
\newblock {\em Schlieren and Shadowgraph Techniques: Visualizing Phenomena in
  Transparent Media}.
\newblock Springer-Verlag, 2001.

\bibitem{toepler1864beobachtungen}
A.~Toepler.
\newblock {\em Beobachtungen nach einer neuen optischen Methode: e. Beitr. zur
  Experimental-Physik}.
\newblock Max Cohen \& Son, Bonn, Germany, 1864.

\bibitem{kokkos}
C.~Trott, L.~Berger-Vergiat, D.~Poliakoff, S.~Rajamanickam, D.~Lebrun-Grandie,
  J.~Madsen, N.~Al~Awar, M.~Gligoric, G.~Shipman, and G.~Womeldorff.
\newblock The {Kokkos} {Ecosystem}: {Comprehensive} {Performance} {Portability}
  for {High} {Performance} {Computing}.
\newblock {\em Computing in Science Engineering}, 23(5), 2021.

\bibitem{RQS}
I.~Wald, M.~Jaros, and S.~Zellmann.
\newblock {Data} {Parallel} {Multi‐GPU} {Path} {Tracing} using {Ray} {Queue}
  {Cycling}.
\newblock {\em Computer Graphics Forum}, 42, 2023.

\bibitem{brix}
I.~Wald and S.~G. Parker.
\newblock {Data Parallel Path Tracing with Object Hierarchies}.
\newblock {\em Proceedings of the ACM on Computer Graphics and Interactive
  Techniques}, 5(3), 2022.
\newblock (Proceedings of High Performance Graphics).

\bibitem{vopat}
I.~Wald, S.~Zellmann, and N.~Morrical.
\newblock {Data-Parallel} {Volume} {Path} {Tracing} (with {Ray} {Forwarding}).
\newblock Available at \url{https://github.com/ingowald/vopat},
  Accessed:~2026/03/16.

\bibitem{mpi}
D.~W. Walker and J.~J. Dongarra.
\newblock {MPI:} a standard message passing interface.
\newblock {\em Supercomputer}, 12, 1996.

\bibitem{Yates:1993}
L.~A. Yates.
\newblock Images constructed from computed flowfields.
\newblock {\em AIAA Journal}, 31(10), October 1993.

\end{thebibliography}

\end{document}